 \documentclass[sigconf]{acmart}

\usepackage{subcaption}
\usepackage{longtable}
\usepackage{graphicx}
\usepackage{array}
\usepackage{geometry}
\usepackage{listings}
\usepackage{fancyvrb}
\usepackage{xcolor}
\usepackage{verbatimbox}
\usepackage{fvextra}
\usepackage{multirow}
\usepackage{stfloats} 
\setcopyright{cc}
\setcctype[4.0]{by}
\DefineVerbatimEnvironment{MyVerbatim}{Verbatim}
  {frame=single,fontsize=\small,breaklines=true,breaksymbolleft={} }
\AtBeginDocument{%
  \providecommand\BibTeX{{%
    \normalfont B\kern-0.5em{\scshape i\kern-0.25em b}\kern-0.8em\TeX}}}

\copyrightyear{2025}
\acmYear{2025}
\setcopyright{cc}
\setcctype{by}
\acmConference[CHI '25]{CHI Conference on Human Factors in Computing Systems}{April 26-May 1, 2025}{Yokohama, Japan}
\acmBooktitle{CHI Conference on Human Factors in Computing Systems (CHI '25), April 26-May 1, 2025, Yokohama, Japan}\acmDOI{10.1145/3706598.3713405}
\acmISBN{979-8-4007-1394-1/25/04}

\begin{document}

\title{eaSEL: Promoting Social-Emotional Learning and Parent-Child Interaction through AI-Mediated Content Consumption
}




\author{Jocelyn Shen}
\authornote{Work done during internship at Apple}
\affiliation{%
  \institution{MIT Media Lab}
  \streetaddress{75 Amherst Street}
  \city{Cambridge}
  \state{MA}
  \country{USA}
}
\email{joceshen@mit.edu}

\author{Jennifer King Chen}
\affiliation{%
  \institution{Apple}
  \city{Seattle}
  \state{WA}
  \country{USA}
}
\email{jennkingchen@apple.com}

\author{Leah Findlater}
\affiliation{%
  \institution{Apple}
  \city{Seattle}
  \state{WA}
  \country{USA}
}
\email{lfindlater@apple.com}

\author{Griffin Dietz Smith}
\affiliation{%
  \institution{Apple}
  \city{Seattle}
  \state{WA}
  \country{USA}
}
\email{griffind@apple.com}

\renewcommand{\shortauthors}{Shen et al.}

\begin{abstract}

As children increasingly consume media on devices, parents look for ways this usage can support learning and growth, especially in domains like social-emotional learning. We introduce eaSEL, a system that (a) integrates social-emotional learning (SEL) curricula into children’s video consumption by generating reflection activities and (b) facilitates parent-child discussions around digital media without requiring co-consumption of videos. We present a technical evaluation of our system’s ability to detect social-emotional moments within a transcript and to generate high-quality SEL-based activities for both children and parents. Through a user study with $N=20$ parent-child dyads, we find that after completing an eaSEL activity, children reflect more on the emotional content of videos. Furthermore, parents find that the tool promotes meaningful active engagement and could scaffold deeper conversations around content. Our work paves directions in how AI can support children’s social-emotional reflection of media and family connections in the digital age.

\end{abstract}



\maketitle

\section{Introduction}
In today's digital age, children aged 5--8 spend on average around three hours per day on devices consuming media, with an overwhelming majority of that time spent watching videos \cite{kabali_exposure_2015, movie__tv_reviews_for_parents_common_nodate}. Unsurprisingly, parents prefer a child's media consumption be beneficial towards their child's learning and emotional needs \cite{neumann_young_2020, neumann_young_2015, smith_contextq_2024}. 

Prior research in digital mediation has shown that children's active engagement with and reflection on the content they consume can yield higher levels of learning in areas including self-efficacy \cite{rasmussen2016relation}, emotion recognition \cite{rasmussen2016relation}, and language development \cite{zimmerman2009teaching}. As such, ample prior works develop solutions that support joint media engagement, in which parents co-consume and, critically, \textit{talk about} that media with their children \cite{smith_contextq_2024, ewin2021impact}. However, joint engagement requires considerable time and attention from the parent, so children often engage independently instead~\cite{movie__tv_reviews_for_parents_common_nodate}. Therefore, we look to facilitate content reflection in an independent use context by supporting children's active learning in-the-moment and by scaffolding parent engagement post-hoc. 

More specifically, parents have directly expressed interest in ways technology can support the development of social-emotional skills \cite{smith_contextq_2024}, a key aspect of children's development. Successful SEL curricula typically involve activities the child should carry out beyond the classroom and suggest active family involvement in the child's social-emotional learning \cite{slovak_teaching_2015}. Consequently, a practical at-home solution for promoting a child's social-emotional growth should involve two parts: (1) facilitating moments of practice outside the classroom and (2) scaffolding parental support and engagement with SEL curricula. 

Aligned with these two lessons from SEL curricula and keeping independent use contexts at the forefront, we developed and evaluated eaSEL (\textbf{e}ducational \textbf{a}ctivities for \textbf{s}ocial-\textbf{e}motional \textbf{l}earning), a prototype system for children aged 5--8 that facilitates independent active engagement with SEL topics during video watching sessions and provides conversation-starters for parents to engage with the child post-hoc without consuming the content themselves. 
Our system leverages large language models (LLMs) to: (1) detect moments for SEL teaching in children's video content, (2) generate SEL activities for children (e.g., drawing a picture or taking a video), and (3) generate parent-child conversation starters. We benchmark the system's performance in conducting these tasks through human evaluation metrics. 
This evaluation reveals that language models can effectively detect moments where social-emotional skills occur, and can generate suitable child-focused SEL activities as well as parent-oriented conversation starters, but fall short in a few areas such as social skill detection and using child-appropriate language.

Finally, we conducted a within-subjects user study with $N=20$ parent-child dyads to assess how eaSEL impacts children's active engagement with SEL and to gauge parent perceptions of the conversation starters. 
Findings show that watching a video with an eaSEL activity results in children using more emotion words in story retellings than when they do not do any activity. Child-produced artifacts from the activities and parent interview data also point to children's SEL engagement. Finally, parents overwhelmingly find that the tool promotes meaningful reflection during otherwise passive consumption. 

In summary, this paper contributes:
\begin{enumerate}
    \item The design and implementation of eaSEL, a system that (1) generates SEL activities alongside a child's media consumption and (2) supports discussion between parents and children around SEL lessons present in media without necessitating co-viewing.
    \item A technical evaluation using human evaluation metrics to assess language model's capacities to identify SEL teachable moments and generate relevant learning activities for both children and parents from those moments.
    \item Empirical evidence of eaSEL's impact on children's SEL engagement and potential parent-child interactions. 
\end{enumerate}

\section{Background and Related Work}

\subsection{Social-Emotional Learning Curricula}
Decades of research indicate that social-emotional skills, such as emotion regulation and empathy, are crucial for a child's academic success \cite{denham_plays_2010} and long-term well-being \cite{zins_social_2007}. In the last 30 years, SEL curricula have been deployed in thousands of schools around the United States \cite{durlak_what_2022, bridgeland_missing_2013, payton_social_2000}. These programs are designed for K--12 learners and follow a framework that includes five core competencies introduced by CASEL (the Collaborative for Academic, Social, and Emotional Learning): (1) self-awareness, (2) self-management, (3) social awareness, (4) relationship skills, and (5) responsible decision making \cite{elbertson_school-based_2009, noauthor_safe_nodate}. Instructional methods for practicing these competencies in the classroom include role play or watching related videos. Teachers also recognize the importance of SEL practice outside of the classroom and often assign homework \cite{elias_promoting_1997}, such as worksheets for emotion identification or reflective writing exercises.

However, a child's social-emotional development does not happen in a formal learning vacuum. From birth, children look to their parents as role models for social behavior \cite{brumariu_mother-child_2015} and face-to-face interactions through different contexts, including childcare or households, provide microsystems that inform a child's socialization \cite{belsky_determinants_1984, li_how_2023, yucel_siblings_2015, dunn_family_1991}. Consequently, SEL educators also develop programs geared towards family involvement, most commonly creating workshops where parents can learn about the child's curriculum \cite{darling_social_2019, mccormick_effects_2016}. However, such workshops have low uptake due to parents' busy schedules or lack of interest in their child's learning \cite{slovak_scaffolding_2016}. As such, these works highlight both the importance and challenges of (1) facilitating independent SEL practice for children in relevant ways and (2) incorporating parents in their children's learning process.

\subsection{HCI for Social Emotional Learning}

Prior work in HCI indicates that it is possible to create digital media interactions that can bolster social-emotional skills. 
Research on fostering social-emotional development for children typically focuses on emotion regulation and processing~\cite{santos_therapist_2020, speer_mindfulnest_2021, theofanopoulou_exploring_2022, slovak_designing_2023, sadka_interactive_2022, buckmayer_storytelling_2024} through means like storytelling \cite{zhou_my_2024} and tangible interfaces \cite{speer_mindfulnest_2021}, although some work has explored relationship skills, such as mediation during conflict resolution \cite{shen_stop_2018}. One early example is the SEL Transition Wheel, an artifact to help children ages 3-6 develop emotion labeling skills \cite{stangl_sel_2017}. More recent works draw on LLMs to design chatbots geared for children's socioemotional competence. For example, \citet{seo_chacha_2024} designed \textit{ChaCha}, a conversation agent that guides children in personal experience sharing and emotion processing. Similarly, \citet{fu_self-talk_2023} presented an interactive conversational AI to deliver SEL lessons, increasing the recall of these lessons post-intervention. Finally, \citet{tang_emoeden_2024} introduce \textit{EmoEden}, which uses text-to-image models and conversational interaction to support emotion learning for autistic children.

However, children already spend a significant amount of time consuming content, and technology could additionally serve the role of augmenting \textit{existing} engagement with active learning. Efforts to modify existing consumption for adults have included embedding wellness interventions into daily Facebook interactions ~\cite{munson_happier_2010} and explorations into human-AI interactions that promote critical thinking during YouTube videos or when reading impact statements for scientific papers \cite{tanprasert_debate_2024, mukherjee_impactbot_2023}. These works, however, did not focus on child participants or pedagogical goals.

An in-situ approach is further supported by research from the learning sciences demonstrating that children are better able to learn from TV shows when learning exercises are aligned with the content of the shows \cite{fisch_childrens_2004, hirsh-pasek_putting_nodate}. As such, it is important to develop systems that provide active learning interactions in-situ with content children are consuming. Most HCI interfaces for children's SEL focus on emotion regulation (as the outcome) and conversational agents (as the methodology), but in our work, we evaluate how LLMs can be leveraged to generate learning activities from a diverse range of children's videos with broad coverage of social-emotional competencies.

\subsection{Digitally-Mediated Parent-Child Interaction}

Prior work on the role of parents in mediating a child's digital usage largely centers on TV-watching, showing that active mediation (i.e., instructive content-related discussion) during co-viewing can support healthy media usage and improve learning outcomes \cite{clark_parental_2011, valkenburg_developing_1999, jiow_level_2017}. Joint media engagement extends the active mediation and co-viewing notion beyond television to encompass the various forms of media families might engage with together \cite{takeuchi2011new}.

Interestingly, joint media engagement studies have found that parents who engage in \textit{active, emotional mediation}, discussing the feelings evoked about media and how to manage those feelings, resulted in their children demonstrating higher emotional intelligence and empathy \cite{nabi_does_2022}. Numerous systems in HCI literature specifically focus on joint media engagement to support learning outcomes \cite{shin_designing_2021, cassidy_cuddling_2024, ho_its_2024, vezzoli_exploring_2020, yu_parent-child_2024, troseth_enhanced_2019, krcmar_parentchild_2014}. For example, \citet{zhang_storybuddy_2022} introduce \textit{StoryBuddy}, an AI-driven system for generating questions parents can ask children during reading, and \citet{kwon_captivate_2022} designed \textsc{Captivate!}, a play-based contextual language guidance system for parents and children from immigrant families. 
However, few works around joint media engagement are tailored for a child's social-emotional development. One prior design study used technology probes to understand considerations in scaffolding parent-child interaction geared for social emotional learning \cite{slovak_scaffolding_2016}. They found that scaffolding behaviors when the parents shifted attention from the device to one another transformed interactions into joint discussions around the digital content. Separately, \citet{smith_contextq_2024} developed \textit{ContextQ}, a system which automatically generates pedagogically guided questions during parent-child storybook reading. They found that families preferred generated questions that covered SEL topics (i.e., questions tying back to the child's life, emotions, or moral lessons), despite SEL not being the focus of this work.

While joint media engagement interactions in HCI have shown empirical promise in learning outcomes and promoting the parent-child bond, parents do not always have the time to co-consume media with their children \cite{shin_designing_2021, radesky_use_2016}. Furthermore, systems that focus on joint media engagement contexts are typically geared towards the parent directing questions at the child, which can reduce parental engagement if the child's learning is irrelevant to the parents' own interests \cite{theofanopoulou_they_2024}.
In our work, we explore how to automatically generate parent conversation starters that give space for the parent and child to mutually share personal experiences and SEL learning.

\begin{table*}[t!]
\footnotesize
\begin{tabular}{ p{.15\textwidth}  p{.2\textwidth} p{.25\textwidth} p{.25\textwidth} } 
\hline \textbf{Area} & \textbf{Example Skill} & \textbf{Positive Example} & \textbf{Negative Example }

\\ \hline Relationship skills & Demonstrate social skills such as helping, giving compliments, and apologizing &  
Max is angry at Joe for stealing 
his ice cream. Joe comes over 
to Max to apologize, and buys 
him new ice cream.
 &  
Max is angry at Joe for stealing his 
ice cream. Joe is full of pride and 
says "well, you should have eaten it 
faster"
 
\\ \hline  
Self-awareness
 & Identifying one's own feelings &  
Max sees Joe walk away, and his chest pangs. "I must be sad that Joe is leaving," Max thinks.
 &  
Max sees Joe walk away, and his chest pangs. Max feels irritated and snaps at a nearby dragonfly, instead of recognizing his sadness.

\\ \hline  
Self-management
 &  
Displaying grit, determination or 
perseverance
 &  
Max is almost at the finish line. 
His chest hurts from running, but 
he thinks to himself, "I can make 
it!" Max finishes in record- 
breaking time.
 &  
Max is almost at the finish line. His 
chest hurts from running, so he thinks 
"I can't do it..." Max gives up and lies 
down on the ground.
 
\\ \hline  
Social awareness
 &  
Perspective taking/empathy
 &  
Max notices that Joe never has enough to eat during lunch. Max decides to share his lunch with Joe everyday.
 &  
Max notices that Joe never has enough to eat during lunch. He rolls his eyes and thinks "who cares?"

\\ \hline  
Responsible decision
making
 &  
Make decisions based on moral, 
personal, or ethical standards
 &  
Max is about to steal Joe's 
lunch, but then he thinks 
"Stealing is wrong."
 &  
Max is about to steal Joe's lunch, but 
then he thinks "Who cares if I steal? 
Joe can just buy another lunch."
 \\
\hline

\end{tabular}

\caption{Examples of social-emotional learning skills in our system pulled from the CASEL framework. Positive and negative examples show demonstration or failure to demonstrate a competency, and are used throughout our activity generation prompting pipeline.}
    \label{SEL-skills}
    \Description{Table with 4 columns (area, skill, positive example, negative example) and 5 rows for each type of social-emotional learning skill (relationship skills, self-awareness, self-management, social awareness, and responsible decision making)}
    \end{table*}

\section{eaSEL Design and Implementation} 
We designed and implemented eaSEL to (1) enhance a child's reflection and active learning during independent media consumption and (2) facilitate moments for the parent to engage with the child's reflection and learning in personally relevant and engaging ways \textit{without} requiring joint media engagement.

\subsection{SEL Learning Objectives} \label{SEL-skillset}
We look to relevant SEL literature to identify target skills for practice. The CASEL framework identifies five key SEL areas to focus on (Table \ref{SEL-skills} lists the areas and examples), and we draw on a set of 10 more specific skills identified as core components of these five areas \cite{lawson_core_2019}. We further develop this skillset with definitions from resources created by SEL educators
\cite{slovak_designing_2015, bar-on_educating_2007, saarni_development_1999, jones_navigating_nodate} and list the final set of SEL skills in the Supplementary Materials.

We also look to classroom activities to understand how educators facilitate practice of these skills. Educators aim to support critical thinking and reflection of social-emotional scenarios \cite{waajid_infusing_2013} and we identify four interactive activities that they use to achieve these goals \cite{slovak_teaching_2015}: 
\begin{enumerate}
    \item \textbf{Drawing and describing personal experiences.} Drawing activities can enhance children's social-emotional development \cite{zakaria_drawing_2021}. Through drawing, children are able to exercise their ability to aesthetically express their identities, fostering self-awareness and social awareness.
    \item \textbf{Changing stories with creative story play.} Creative storytelling and imaginative play can foster a child's social-emotional development by allowing children to explore alternative outcomes across social scenarios \cite{nance_mindful_2018, uslu_improving_2021}, aiding in decision making and predicting others' feelings.
    \item  \textbf{Telling personal stories.} Self-reflection exercises are core to many SEL curricula, as such reflections allow students to transfer knowledge of social scenarios into their own life \cite{waajid_infusing_2013}. 
    \item \textbf{Acting or role playing scenarios.}  Role-playing allows students to embody the experiences of others, aiding in the development of relationship skills such as cooperation and empathy \cite{heyward_emotional_2010, vlaicu_importance_2014, zhang_e-drama_2009}. 
\end{enumerate}

\begin{figure*}[t]
    \centering
    \includegraphics[width=\linewidth]{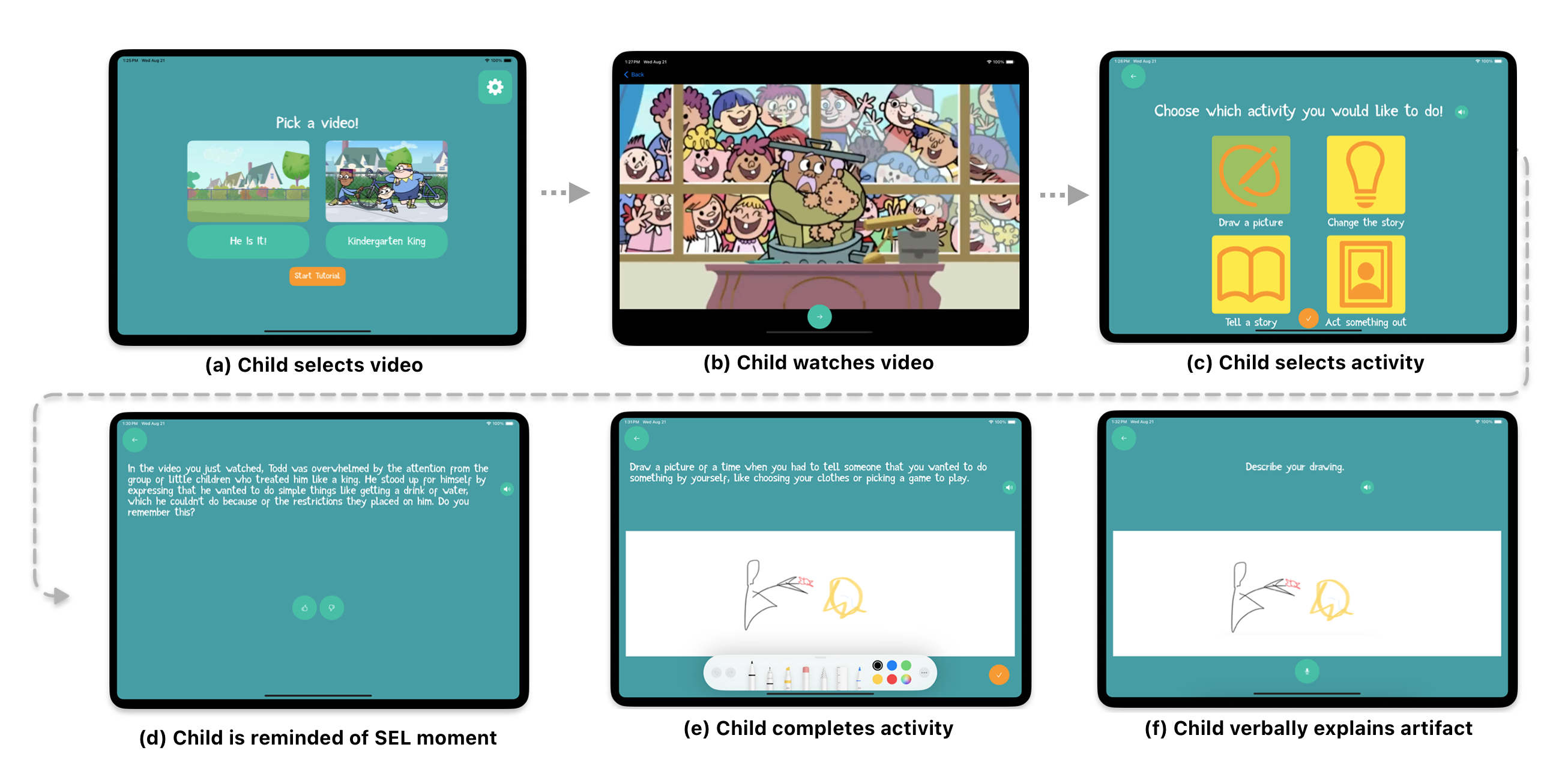}
    \caption{Example app flow for the child user. The interaction starts when children (a) select a video, (b) watch a video, then (c) select an activity. The child is first reminded of the moment where the social-emotional skill occurred in the show (d) and then they complete the activity (e). Finally, the child verbally explains the artifact they created if the activity did not involve audio recording (f).}
    \label{childui}
    \Description{Depicts a step-by-step flow of a child's interaction with an app. It begins with the child selecting a video to watch. After watching the video, the child chooses an activity related to the video's content, such as drawing a picture or changing the story. The child is then reminded of a specific social-emotional learning moment from the video and proceeds to complete the chosen activity. If the activity doesn't involve an audio recording, the child then verbally explains what they created.}
\end{figure*}

\subsection{User Interface Design}
We design a two-part interface grounded in these SEL learning objectives and practices.

\subsubsection{Child Interface}
A child interface (see Figure \ref{childui}) allows children to watch video content and complete video-grounded SEL activities. Figure \ref{childui} shows a sample app flow from a child user's perspective. First, the child selects a video to watch. After watching the video, the child is directed to an activity selection page, where they select one of four activities aligned with the classroom activities identified in Section \ref{SEL-skillset}.
These activities pull from key social-emotional moments in the show and are designed to inspire deeper reflection on the content. Once a child finishes an activity, they are returned to the video selection screen to continue watching.

The activities themselves consist of three parts. We first provide a short reminder of the relevant moment in the show to help children better connect the activity to the content. Then we prompt with the activity itself. Finally, following lessons from prior work \cite{dietz2021storycoder}, in the case of personal experience sharing we present example responses to help cue memories and constrain an otherwise very open-ended recall problem.
An example activity is: ``\textit{In the video you watched, Moo Moo was very determined to build the biggest and most beautiful nest, even though everyone kept telling her she was a cow and not a bird. Can you tell me about a time when you worked really hard to finish something important to you, even if it was difficult or others didn't believe you could do it? For example, maybe you kept trying to tie your shoes by yourself or finish a puzzle even when it was tough.}''

The user interface is designed to be child-friendly with large fonts and buttons as well as text-to-speech to verbalize questions for children who cannot yet read. Children can press and pause the text-to-speech in order to replay the prompts as many times as they need, and activity selection icons read out the type of activity before proceeding. While Figure \ref{childui} shows one example with a drawing activity, our app supports multi-modal inputs for each of the SEL activities, such as video and audio recording.

\subsubsection{Parent Interface}
\begin{figure*}[t]
    \centering
    \includegraphics[width=.65\linewidth]{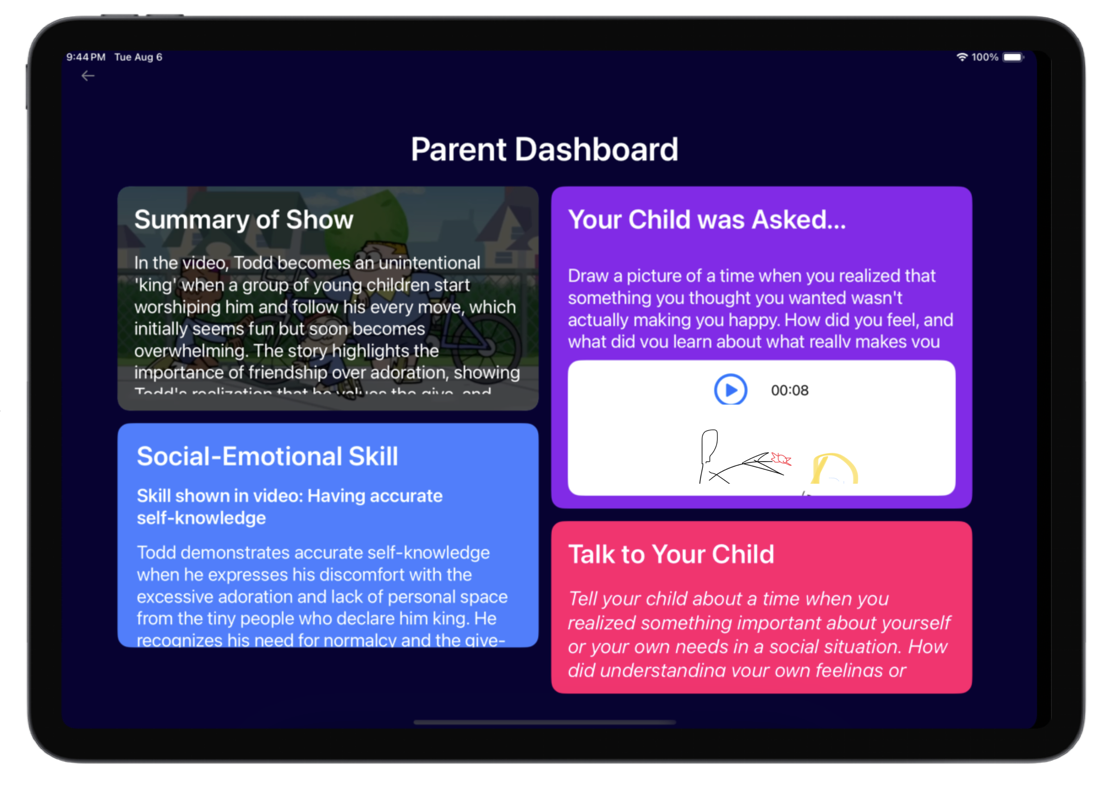}
    \caption{Example parent interface showing a summary of the show, the social emotional skill detected, the child's activity playback, and generated conversation starters.}
    \label{parentui}
    \Description{Once in the parent view, they are presented with an interface that includes a summary of the show their child watched, details about the social-emotional skill addressed in the episode, and conversation starters to discuss the child's activity. The interface also allows parents to view or listen to the child's completed activity.}
\end{figure*}

When a child completes an episode and activity, we populate a parent interface (Figure \ref{parentui}) that presents multiple ways a parent can engage with their child post-hoc. An episode summary reflects existing parent dashboards that are meant to support parent-child discussions over content. We also include an overview of relevant SEL topics to bring that aspect of content to the forefront and show the artifact from the child's activity, including playback of any video or audio recordings the child made; these artifacts are still episode-connected, but are more personally relevant entrypoints for the parent because they reflect the child's thinking and voice. 

Finally, prior literature demonstrates that exchange of personal experiences can support empathetic bonds that promote deeper connections \cite{pillemer_remembering_1992}, that a child's self understanding is often shaped by family stories or narratives shared by parents \cite{miller_narrated_1992, fivush_constructing_1994}, and that sharing of personal stories supports engagement and relevance of the interaction to the parent \cite{stock_telling_2012}. As such, we present a parent-child conversation starter that asks the parent to reflect on their own experiences related to the target SEL skill. Aligned with our objective of supporting independent use contexts, parents can use these prompts to freely engage in discussion with their child at convenient times (e.g., over dinner or before bed), without requiring additional screen time, and without having watched the episode themselves. 

The conversation-starters consist of three parts. First, a sentence explains the SEL moment from the episode to ground the discussion. The next sentence overviews the related SEL skill to provide a more concrete lesson objective for the parent. Finally, a prompt asks the parent to discuss with their child a personal experience related to this SEL skill and moment.
An example conversation starter is: ``\textit{In the video your child watched, Moo Moo shows great determination by deciding to build the biggest and most beautiful nest, even though everyone keeps telling her she's a cow and not a bird. This shows a lot of perseverance. Can you share a story with your child about a time when you had to keep trying hard to achieve something, even though it was difficult or others doubted you?}''

\subsection{Implementation}

We developed and deployed the eaSEL prototype as a tablet-based app in Swift for iOS. The parent user interface is accessed through the app settings page so that children do not accidentally switch to parent view. 

We leverage a pipe-lining approach to generate child activities and parent-child conversation starters. We selected this approach based on extensive prior literature demonstrating that iterative prompting improves generations for complex tasks by breaking down these tasks into smaller parts, compared to a single ``catch-all'' generation prompt \cite{sun_prompt_2024, wu_promptchainer_2022, wang_iteratively_2022, zhou_least--most_2023, arora_ask_2023}. To this end, we first implemented an SEL detection task to identify social-emotional learning moments tied to the identified SEL skills in the video transcripts. Then, using the video transcript and the output of that detection task, we generated activities that promote SEL reflection in children and conversation starters to foster parent-child conversation about the targeted SEL skill. For our implementation, we used Whisper to transcribe videos and OpenAI's GPT-4 ChatCompletion API\footnote{\url{https://openai.com/index/gpt-4/}} for SEL detection and activity generation. Our prompting approach is shown in Figure \ref{fig:chainedprompts} and our full prompts are included in the Supplementary Materials.

In the following sections we present a technical evaluation to benchmark performance on the detection and generation tasks and a mixed methods user study to understand the impact of the developed system on children and their parents.

\begin{figure*}[t!]
    \centering
    \includegraphics[width=\linewidth]{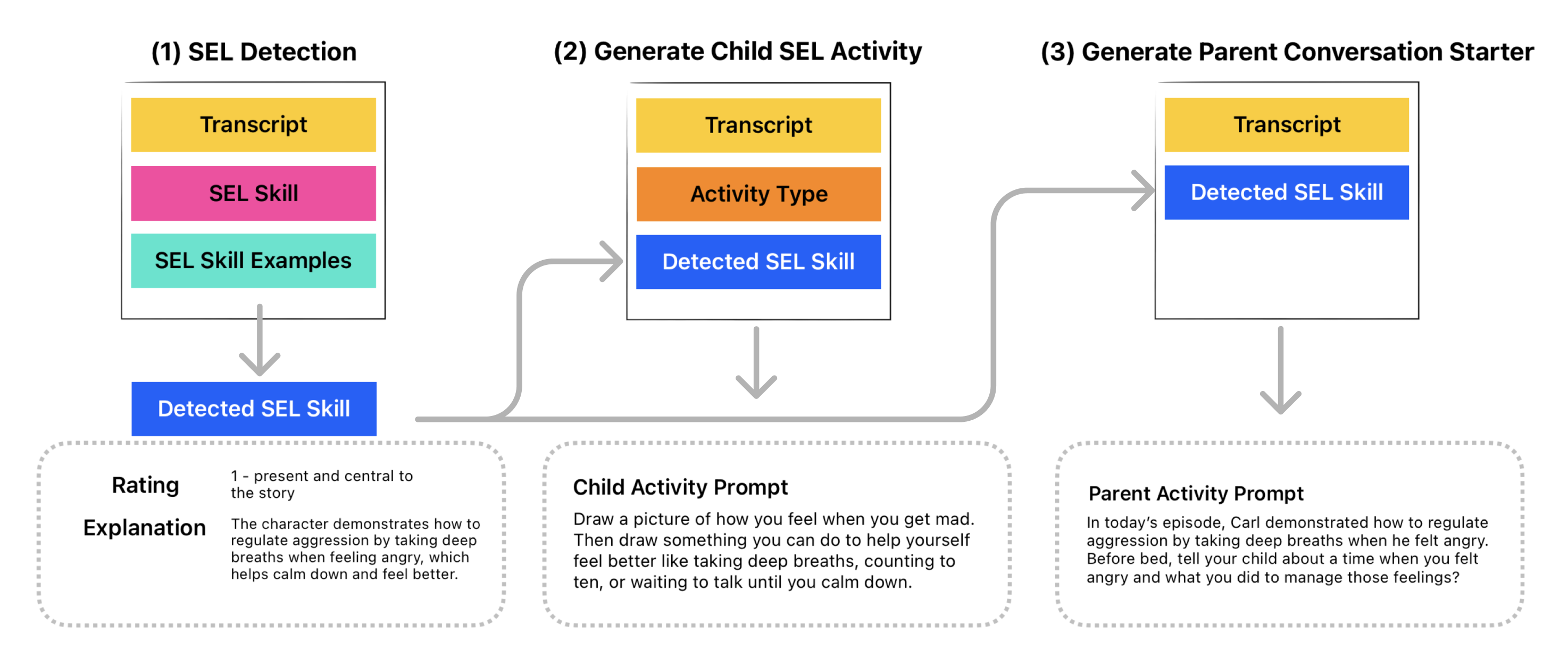}  

    \caption{Overview of pipelined prompting approach, with inputs and example outputs.}
    \Description{Overview of a pipelined prompting approach that includes three main stages: SEL Detection, Generation of Child SEL Activities, and Generation of Parent Conversation Starters. In the first stage, a transcript is analyzed to detect SEL skills, which are then rated and explained. In the second stage, the detected SEL skill is used to generate a child activity prompt based on the transcript and activity type. In the final stage, the detected SEL skill is utilized to create a parent conversation starter related to the episode's content, aiming to encourage further discussion between the parent and child.}
    \label{fig:chainedprompts}
\end{figure*}

\section{Technical Evaluation of Activity Generation}
We benchmark performance of the individual steps of our pipelined activity generation method against human evaluators in order to ensure high performance across these three tasks in a fully deployed system. From this evaluation, we show that LLMs can detect key SEL moments in transcripts and generate suitable activities for both children and parents to engage with social-emotional skills. Our full annotation instructions for all tasks are included in the Supplementary Material.

\subsection{Dataset}
We identified 53 children's TV episodes published online with creative commons licenses by animation studios, and generated automated transcripts of those videos using Whisper \cite{radford_robust_nodate}. All of these videos were tagged as being appropriate for children in our age range (5--8), and ranged from 7 to 20 minutes long. We use 43 of these videos for the evaluation tasks, with the ten remaining videos set aside for iterating on annotation instructions.

\subsection{Identifying SEL Teachable Moments}\label{teachable-moments-task}
In the first task, we use in-context learning with GPT-4 to detect social-emotional events in the transcripts \cite{brown_language_2020}. For each transcript, we loop through each of the 10 SEL skills derived from prior work (Table \ref{SEL-skills}) and prompt for a binary rating of whether the skill is present in a ``main event'' (i.e., a faithful summary of the story could not drop the event) of the story as well as a 1-2 sentence explanation, providing examples of each skill.

\subsubsection{Evaluation Method}
After five iterations of training and alignment using the ten held-out transcripts, two members of the research team independently rated presence of each of the 10 SEL skills in our skill set and wrote 1-2 sentence explanations for their ratings. We calculated inter-rater reliability between the 2 annotators, reporting both  Krippendorff's alpha (KA), a common measure of inter-rater reliability that accounts for random agreement but can be low with unbalanced classes (i.e., more skills marked not present than present for a given video) and overall percent agreement which is less skewed by unbalanced data but does not account for random chance. We achieve a Krippendorff's alpha of 0.64 (substantial agreement) \cite{wong_cross-replication_2021, landis_measurement_1977} and an overall percentage agreement of 0.88. To obtain final gold labels for both ratings and explanations, the two annotators met to discuss and resolve disagreements.

We use the following evaluation metrics to quantitatively assess performance of LLMs on detecting social-emotional events compared to human gold ratings, in addition to qualitatively exploring failure cases:
\begin{itemize}
    \item \textbf{Accuracy} -- For each skill, we analyze the number of correct predictions over total number of predictions.
    \item \textbf{F1 Score} -- F1 score is a measure balancing the precision and recall of the predictions and is a suitable evaluation metric for cases where classes are imbalanced. 
    \item \textbf{Sentence-BERT Cosine Similarity} -- We compute cosine-similarity between Sentence-BERT \cite{reimers_sentence-bert_2019} embeddings of GPT-4-written and human-written explanations of why a social-emotional moment is present in the transcript to understand whether humans and GPT-4 identify similar moments.\footnote{We use the paraphrase-MiniLM-L6-v2 model .} These scores can be computed only for transcripts where both the human and GPT-4 provided such explanations (i.e., identified that skill as present).
\end{itemize}

\subsubsection{Results and Discussion}
Table \ref{detection-results} shows performance of GPT-4 in detecting SEL moments in transcripts. Compared to human ratings, GPT-4 performs well in identifying each of the SEL skills in our skillset, with accuracy and F1 scores over 70\% for all skills except for R2, ``demonstrate social skills''. This weakness of LLMs echoes prior work showing that social reasoning tasks can be challenging for models \cite{sap_neural_2022}, especially when dealing with the interaction between two or more characters \cite{shen_modeling_2023}. Notably, of the samples that GPT-4 incorrectly identified, 24/25 were false negatives. We note that in our system false negatives are less harmful than false positives, as the system selects only one of the detected SEL skills to generate activities for the child; whereas missing a present skill may have no impact, false positives (i.e., identifying a skill that is not present) could introduce more confusion to the child.

\begin{table*}[t!]
    \centering
    \setlength{\extrarowheight}{2pt}
    \renewcommand{\arraystretch}{1.2}
    \resizebox{0.65\textwidth}{!}{%

    \begin{tabular}{c|c|m{4.5cm}|c|c|c}
    \hline
    \textbf{Skill Category} & \textbf{Skill ID} & \textbf{Skill Description} & \textbf{Accuracy} & \textbf{F1} & \textbf{Cosine Similarity of Explanations} \\
    \hline Self-awareness (A) &  A1 & Identifying one's own feelings & 81.40 & 76.55 & 38.80\\
    \hline
    Self-awareness (A) & A2 & Having accurate self-knowledge & 93.02 & 92.37 & 56.48 \\
    \hline
    Self-management (M)& M1 & Regulating negative emotions including impulse, stress, aggression, and pessimism & 76.74 & 71.39 & 32.10\\
    \hline
    Self-management (M)& M2 & Displaying grit, determination, or perseverance & 74.42 & 76.53 & \underline{19.80} \\
    \hline
    Social awareness (S) & S1 & Identifying other people's feelings & 90.70 & 89.26 & 23.19\\
    \hline
    Social awareness (S) & S2 & Perspective taking/empathy & 86.05 & 85.69 & 23.93 \\
    \hline
    Social awareness (S) & S3 & Respecting diversity and different viewpoints & \textbf{97.67} & \textbf{98.05} & \textbf{73.96}\\
    \hline
    Relationship skills (R) & R1 & Standing up for oneself & 90.70 & 90.70 & - \\
    \hline
    Relationship skills (R) & R2 & Demonstrate social skills such as helping, giving compliments, and apologizing & \underline{41.86} & \underline{51.27} & 21.80\\
    \hline
    Responsible decision making (D)& D1 & Making decisions based on moral/ethical standards & 86.05 & 84.48 & 35.41\\
    \hline

    \end{tabular}
    }
    \caption{For each SEL skill, we evaluate how well GPT-4 detects social-emotional moments in transcripts, compared to human gold labels. Note that all scores are multiplied by 100 for easier comparison, and the maximum for each metric is 100. In \textbf{bold} is the best performing and \underline{underlined} is the worst performing condition for the metric.}
    \Description{The table shows how well GPT-4 detects social-emotional learning (SEL) moments in transcripts by comparing its performance to human gold labels. The table presents three metrics for each skill: Accuracy, F1 score, and Cosine Similarity of Explanations, with all values multiplied by 100 for easier comparison. The highest performance for each metric is bolded, while the worst performance is underlined. Notably, the skill "Respecting diversity and different viewpoints" (S3) shows the highest accuracy (97.67), F1 score (98.05), and cosine similarity of explanations (73.96).}
    \label{detection-results}
\end{table*}

On average, cosine similarity between human-written and GPT-written explanations was $0.36$, which shows weak similarity. However, when investigated qualitatively we observe that even low similarity explanations are reasonably similar to one another; GPT-4 outputs tend to be more verbose and refer to character names less frequently, which could decrease cosine similarity. For example, the following human and GPT-4 explanations had a cosine similarity of only 0.18, but both described one character helping another find their way out of the forest: for the human explanation 
 ``\textit{Froggy offers to help Moo Moo get out of the forest. And says they can work together. He reassures her that everything will turn out fine,}'' GPT-4 responded with ``\textit{The transcript shows a positive example of social skills when the character offers to help another character who is lost in the forest. This is a central plot point as it revolves around the main conflict of the character being lost and needing assistance to find their way back, demonstrating cooperation and helping behavior.}'' 
Overall, many pairs have substantial semantic similarity, and even lower similarity pairs refer to similar reasons.

\subsection{Generating Activities for Children}
In the next task we generate a reflection activity from a show transcript, a specified activity type (see Section \ref{SEL-skillset}), and the identified SEL skill/moment from the detection task. We again use in-context learning for this generation task, providing examples of suitable activities.

For this evaluation, we draw on \citet{smith_contextq_2024}'s rubric for assessing generated questions for children in addition to prior work in social-emotional learning \cite{slovak_teaching_2015}, to identify suitability criteria across three categories: syntax/structure of language, relevance, and promotion of SEL. Specifically, the activity prompt must be worded appropriately for a young audience \cite{charlesworth_developmentally_1998}, must be relevant to the SEL skill and the moment it appears in the story \cite{fisch_childrens_2004, hirsh-pasek_putting_nodate}, and should encourage critical engagement and reflection \cite{elias_promoting_1997}. As no automatic evaluation metrics exist for this task, we rely on human annotators to assess whether generated activities for children are suitable across a variety of metrics.

\subsubsection{Evaluation Method}
For our human evaluation, we recruited five annotators who were native English speakers and parents. 
For each TV episodes in our dataset and each SEL skill present in that episode (determined by the previous detection task), we randomly selected one of the four activity types. From these selections (episode, skill, and activity type), we generated a total of 59 child activities. For each activity, we obtained annotations from every annotator across all evaluation metrics. To evaluate the quality of generated activities, we use the following human evaluation metrics:
\begin{itemize}
    \item \textbf{SEL skill relevance} -- The generated activity aligns with the detected skill (Likert 1-5).
    \item \textbf{SEL moment relevance} -- Activities generated are related to the detected SEL moment rather than a spurious part of the story (Likert 1-5).
    \item \textbf{Activity-grounded} -- The question prompts the child to do the specified activity type (Binary yes/no).
    \item \textbf{Child-appropriateness} -- The generated activity is developmentally appropriate for children 5--8 years old based on the following aspects (all Likert 1-5) \cite{saywitz_facilitating_1999, saywitz_credibility_1993, imhoff_preschoolers_1999}:
    \begin{itemize}
    \item Lexical simplicity: sentences do not contain complex words (i.e., more than 3--4 syllables or technical jargon).
    \item Syntactic simplicity: sentences are not too long or complex.
    \item Topic shifts: sentences do not rapidly or frequently shift in topic.
    \item Topic familiarity: sentences contain topics that are familiar to children ages 5--8.
     \end{itemize}
    \item \textbf{Other question suitability metrics} (Binary yes/no):
    \begin{itemize}
    \item The activity does not contain nested questions.
    \item The activity is not a yes or no question.
    \item The activity does not involve another person.
    \end{itemize}
    \item \textbf{Reflection} -- The activity promotes reflection by satisfying at least one of the six possible reflection criteria (checklist): (1) Reflection usually relates to an experience, (2) May involve acknowledging and examining feelings, (3) Provides the opportunity to view different perspectives, (4) Increases self-awareness, (5) Provides a basis for change, and (6) Considers alternative actions \cite{gibbs_learning_1988,thompson_reflective_2021}.
\end{itemize}

We use a Mean Opinion Score (MOS) test \cite{viswanathan_measuring_2005}, a common method in speech, image, and text quality evaluation that relies on computing means of multiple opinions from different annotators across model output samples. We additionally report percentages of responses that fall within Likert categories for the metrics above, reflecting quality assessment methods from NLP summarization work \cite{zhao_narrasum_2022}. Finally, we qualitatively explore open-ended responses from annotators for cases when they do not believe an activity is suitable.

\begin{figure*}[t!]
    \centering
    \includegraphics[width=.9\linewidth]{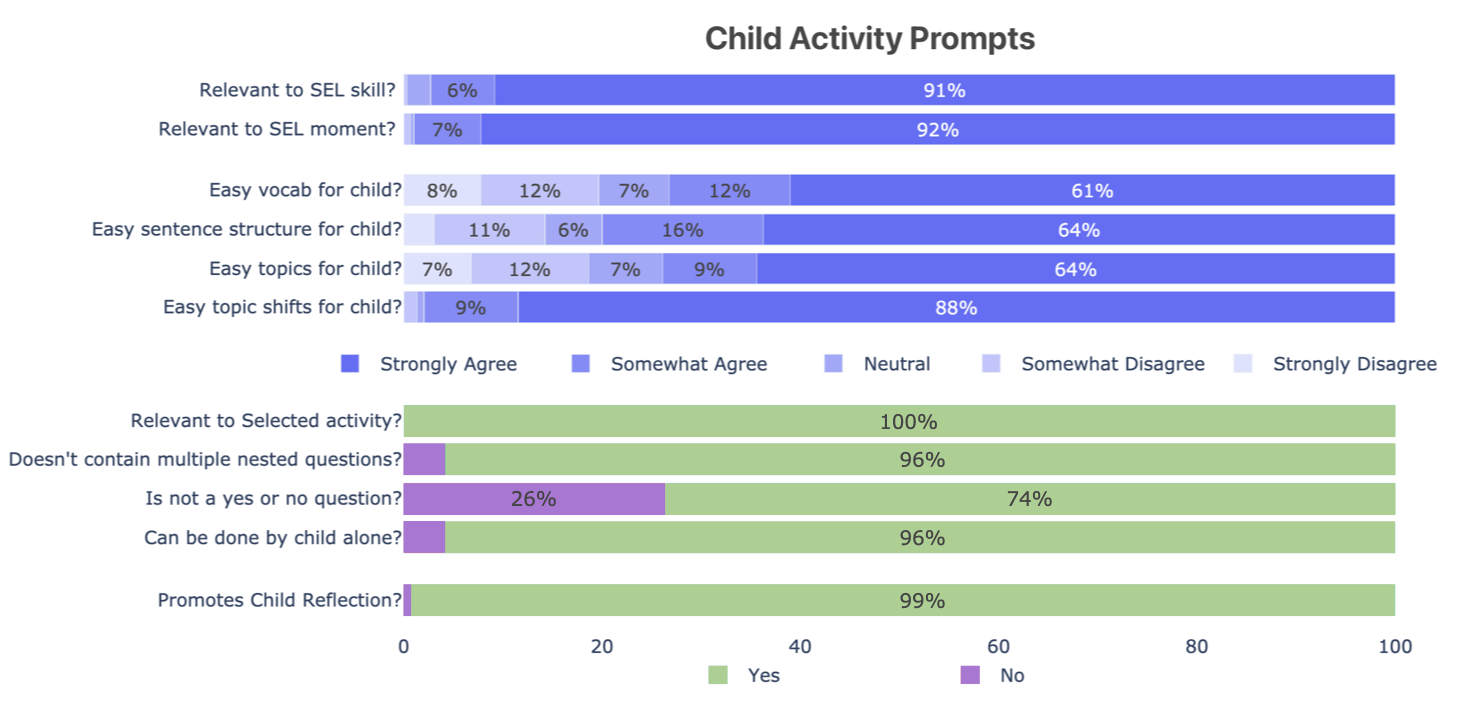}
    \caption{Human evaluation of GPT-4 generated child-focused SEL activities, with five annotators for each of the 59 generated child activities.}
    \Description{The figure presents a human evaluation of GPT-4-generated child-focused SEL (Social-Emotional Learning) activities. It shows the percentage of responses in categories such as "Strongly Agree," "Somewhat Agree," "Neutral," "Somewhat Disagree," and "Strongly Disagree" for different metrics. Most respondents "Strongly Agree" that the activities are relevant to SEL skills (91\%) and SEL moments (92\%). The activities are also generally considered easy for children, with high agreement on vocabulary, sentence structure, and topic shifts. Other metrics, such as relevance to the selected activity, lack of multiple nested questions, and promoting child reflection, also received high positive responses, with 100\% agreement on relevance and 99\% on promoting reflection.}
    \label{gen_results}
\end{figure*}

\subsubsection{Results and Discussion}\label{tech_eval_child_gen_results}
Figure \ref{gen_results} shows the distribution of annotator responses across all evaluation metrics. Overall, annotators strongly agree that child activities are relevant to the SEL skill and moment (> 90\% of samples with ``Strongly Agree'', and $M > 4.5$). Furthermore, 100\% of the generated activities are aligned with the activity type, and 96\% do not contain nested questions and can be done by the child alone. While the majority of samples do not contain yes or no questions (74\%), this percentage is lower than other categories in the suitability evaluation metrics. Although the presence of yes/no questions may lead to shorter responses, it was not harmful to move forward with them present. 

Annotators also believe the activities support reflection; 99\% of generated activities have one or more reflection criteria. Involving the child in acknowledging or examining feelings was the most prevalent reflection criteria (24.08\% of the samples), while relating to the child's personal experiences was the least prevalent criteria (16.73\% of the samples). 

Critically, our MOS test shows child-appropriate language metrics have lower means than the other categories ($4.10 \leq M \leq 4.90$). While most samples have easy topic shifts for a child (88\% at ``Strongly agree''), fewer examples contain easy vocabulary (61\%), sentence structure (64\%), and topics that are understandable to a 5-8 year old child (64\%). 

Annotators' open-ended explanations pointed to two key areas of concern. First,annotators concerned with language primarily identified specific words they felt might be difficult for a child to grasp (e.g., ``determination,'' ``mystery,'' ``succeeded''). Despite our prompts including specific ways to make outputs child-friendly, these results are somewhat intuitive, as the majority of training data available to language models is not sourced from children's text. For this application, though, if vocabulary is too difficult it could hinder the goal of the activity itself. Conversely, a key goal of media reflection is building vocabulary \cite{rice_lessons_1988} and given the context of the video, the introduction of these new words might present an opportunity for learning. Therefore, we elected to proceed and evaluate comprehension directly with child participants.

Second, annotators occasionally deemed the activity topics themselves as inappropriate. In response to the activity, ``\textit{In the video you watched, Todd stood up for himself by telling Jenny and Mauricio that he did not want them to continue their actions and that he was not interested in a romantic relationship with either of them. Draw a picture of a time when you had to tell someone that you didn't like what they were doing or that something they were doing was making you uncomfortable. For example, if someone was being too loud while you were trying to read, or if someone was playing too roughly and you wanted them to stop
},'' an annotator wrote ``\textit{Five-year-olds should not be exposed to sexualized ideas like dating. This seems very dangerous.}'' These themes, however, come directly from the video content itself. Despite all videos in our dataset being tagged as kid friendly for children aged 5--8, some videos contain themes parents could find inappropriate. We do not explore content moderation in this work so consider the concern out of scope, but future work might explore moderation of content explicitly tagged child-appropriate and whether educational activities should be generated for potentially inappropriate content. 

Overall, we show that our system generates suitable activities for children across our defined metrics, with lower performance in child appropriateness than other metrics, and we carefully select appropriate video content for children to watch in our subsequent user study.

\begin{figure*}[t!]
    \centering
    \includegraphics[width=.9\linewidth]{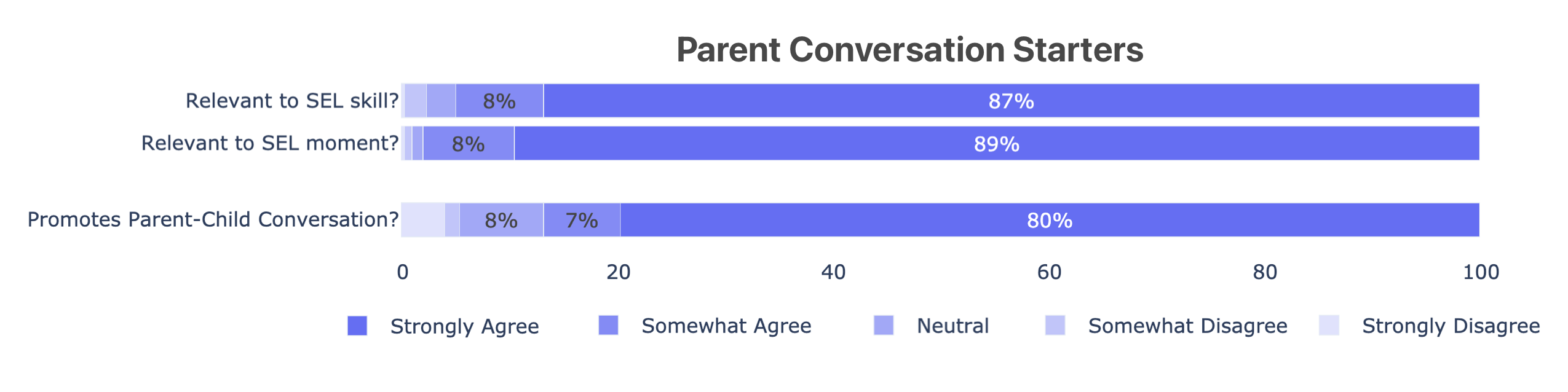}
    \caption{Human evaluation of GPT-4 generated parent-child conversation starters.}
    \Description{The figure shows the human evaluation of GPT-4-generated parent-child conversation starters. It measures how relevant these prompts are to SEL skills, SEL moments, and whether they promote parent-child conversation. The majority of respondents "Strongly Agree" that the prompts are relevant to SEL skills (87\%) and SEL moments (89\%). Additionally, 80\% "Strongly Agree" that the prompts effectively promote parent-child conversation, with smaller percentages somewhat agreeing or being neutral on these metrics.
}
    \label{gen_results_parent}
\end{figure*}

\subsection{Generating Conversation Starters for Parent-Child Interaction}\label{parent-child-task}
We again use in-context learning to generate conversation starters for parents based on the SEL moments in videos.

\subsubsection{Evaluation Method}
The same five human evaluators from the previous task evaluated generated parent conversation starters for the 59 video/SEL skill pairs. We designed the following human evaluation metrics to assess whether LLM-generated activities could promote deeper parent-child conversations around the SEL themes and content. The evaluation metrics are as follows:

\begin{itemize}
    \item \textbf{SEL skill relevance} -- The generated conversation starter aligns with the detected skill (Likert 1-5).
    \item \textbf{SEL moment relevance} -- The generated conversation starter is relevant to the teachable moment in the video the child watched (Likert 1-5).
    \item \textbf{Meaningful dialogue} -- The conversation starter is likely to foster meaningful dialogue between a parent and child, following the survey questions of \citet{smith_contextq_2024}.
\end{itemize}

We repeat the same MOS test and an analysis of annotator response distributions from the child activity generation evaluation. We further qualitatively explore open-ended feedback to explain quantitative performance. 

\subsubsection{Results and Discussion}
Figure \ref{gen_results_parent} shows that annotators ``strongly agree'' that at least 80\% of the generated parent-child conversation starters are relevant to the SEL skill and moment, and that they could promote meaningful parent-child conversation. Relevance to the SEL moment ($M = 4.90$, 95\% CI [$4.862$, $4.938$]), relevance to the detected SEL skill ($M = 4.90$, 95\% CI [$4.853$, $4.947$], and potential to promote parent-child conversation ($M = 4.6$, 95\% CI [$4.515$, $4.685$]) are high on average, with means all over 4.5 (4 = somewhat agree, 5 = strongly agree).

Exploring open-ended explanations, we saw two reasons why annotators found some prompts did not promote meaningful parent-child discussion. First, mirroring the findings from the child activity generation, annotators occasionally disagreed with the appropriateness of the source videos themselves (see discussion in Section \ref{tech_eval_child_gen_results}). Second, a few annotators mention lack of specificity in the prompt. For example, one annotator mentioned, ``\textit{I think it is important to arm the parents with a little bit more than just these very general descriptions. Simple things like the names of the characters are important for the parent to be aware of.}'' While the intention of our generated prompts in connecting to the parent's personal experiences was to facilitate conversation without co-viewing, some annotators felt a little more information about what the child watched would lead to more fruitful discussion. In our system implementation and user study, we include a show summary alongside prompts to understand how parents feel about the time/utility trade off of receiving this additional information.

\subsection{Summary of Findings and System Limitations} \label{systemlimitations}
Overall, we demonstrate that our system can sufficiently detect SEL skills and moments in video transcripts. Generated child activities are relevant to these detected skills and promote reflection, while being broadly appropriate for child users when the source content itself is appropriate. Parent annotators also find generated conversation starters to be both relevant and likely to promote meaningful dialogue between themself and their child, although occasionally felt additional detail on the original content might be useful.

Critically, in our detection task, the system shows inconsistent performance in detecting the R2 social skills category, and we saw disparities in the cosine similarity of LLM vs human explanations, with LLMs often favoring verbosity. Future work in model training may aim to incorporate tasks specific to social reasoning benchmarks as training objectives \cite{sap_socialiqa_2019} or focus on better aligning human and language model social-emotional reasoning explanations (e.g., via fine tuning on human rationales \cite{kabir_beyond_2024}). 

For the generated activities, the model occasionally outputs language that is too advanced for the target age group and some parent annotators raised concerns about generating activities for source content that included potentially inappropriate themes. In typical joint media engagement contexts, parents may intervene or change the topics of discussion when they find content objectionable, but without the parent present to do that content moderation, we face new questions about whether there are types of media for which reflection activities should \textit{not} be generated. Our work reveals that there is a need for future work intersecting content safety moderation approaches \cite{ghosh_circle_2020, saldias_designing_2024, ghosh_safety_2018} as well as better guardrails tailored for child users. Future work could develop child-centric language models fine-tuned on children's content and validated by child development experts, and in the content moderation space, future work might explore integrated automatic content filtering or personalization to what parents deem appropriate.

Moving forward in our user study we carefully select appropriate content and pay specific mind to children's activity comprehension and parent perceptions of the show summary information.

\section{User Study Method} \label{user_study}
While the previous section provides technical evaluation of our generated child and parent activities, we also wanted to understand how families would use and perceive the system. To this end, we conducted a mixed-methods study with $N=20$ parent-child dyads to answer the following research questions:
\begin{itemize}
    \item \textbf{[RQ1]} How does eaSEL impact children's SEL reflection during independent media consumption? \item \textbf{[RQ2]}: How do parents perceive the potential utility or impact of such a system?
\end{itemize}

\subsection{Study Design} \label{sec:study_design}
The study included two parts. The first part focused on \textbf{RQ1} and consisted of a one-way within subjects experiment where the child participants used different versions of the system. This study design included a single experimental factor of Activity that had two levels: 1) \textbf{\textit{No Activity}}, where the child simply watches the video and 2) \textbf{\textit{eaSEL Activity}}, where the child selects and completes an automatically generated activity related to social-emotional learning. The orders of the two episodes and of conditions were counterbalanced across participants. For this part of the study, we quantitatively test the following hypothesis: \textbf{[H1]} Children who complete an \textit{eaSEL Activity} will use more emotional language in  re-tellings of content they consumed compared to doing \textit{No Activity}. We also qualitatively describe the artifacts children produced during the eaSEL activities to understand how they engaged with the presented SEL concepts.

To answer \textbf{RQ2} in the second part of the study, we presented parents with the parent interface and conducted a semi-structured interview based on the interface and ideas underpinning the system. 

\begin{table*}[t!]
    \centering
    \resizebox{.9\textwidth}{!}{
    \begin{tabular}{c|c|c|c|c|c|c|c}
    \hline
    \begin{tabular}{l} 
    Family \\
    ID
    \end{tabular} & \begin{tabular}{l} 
    Child \\
    Age
    \end{tabular} & \begin{tabular}{l} 
    Child \\
    Gender
    \end{tabular} & \begin{tabular}{l} 
    Parent \\
    Age
    \end{tabular} & \begin{tabular}{l} 
    Parent \\
    Gender
    \end{tabular} & \begin{tabular}{l} 
    SEL at \\
    School
    \end{tabular} & \begin{tabular}{l} 
    Conversation about \\
    Digital Media
    \end{tabular} & \begin{tabular}{l} 
    Joint-Media \\
    Engagement Frequency
    \end{tabular} \\
    \hline \hline
    P05, C05 & 8 & Female & 39 & Male & Unsure & Sometimes & Sometimes \\
    P06, C06 & 8 & Female & 39 & Male & Unsure & Often & Often \\
    P07, C07 & 8 & Male & 41 & Female & Unsure & Sometimes & Always \\
    P08, C08 & 7 & Male & 37 & Male & No & Sometimes & Sometimes \\
    P10, C10 & 8 & Female & 47 & Male & Yes & Sometimes & Often \\
    P11, C11 & 6 & Male & 34 & Female & Yes & Always & Always \\
    P13, C13 & 6 & Male & 40 & Male & Unsure & Often & Often \\
    P14, C14 & 7 & Non-binary & 40 & Female & Yes & Often & Sometimes \\
    P18, C18 & 8 & Female & 40 & Female & Yes & Sometimes & Always \\
    P20, C20 & 7 & Female & 45 & Male & Yes & Often & Often \\
    P21, C21 & 7 & Female & 43 & Male & Yes & Sometimes & Often \\
    P24, C24 & 7 & Male & 35 & Male & Unsure & Often & Always \\
    P25, C25 & 5 & Female & 34 & Female & Yes & Often & Sometimes \\
    P26, C26 & 8 & Female & 44 & Male & Yes & Often & Often \\
    P27, C27 & 6 & Male & 45 & Female & Yes & Always & Always \\
    P28, C28 & 8 & Female & 42 & Female & Yes & Sometimes & Rarely \\
    P31, C31 & 6 & Male & 44 & Male & Yes & Often & Often \\
    P33, C33 & 6 & Male & 44 & Male & Yes & Always & Sometimes \\
    P38, C38 & 8 & Female & 35 & Female & No & Sometimes & Sometimes \\
    P39, C39 & 7 & Male & 42 & Male & No & Often & Often \\
    \hline
    \end{tabular}
    }
    
    \caption{Participant demographic information including conversation about digital media and joint-media engagement frequency}
    \Description{The table provides demographic information for study participants, including family ID, child age and gender, parent age and gender, whether the child receives SEL (Social-Emotional Learning) at school, the frequency of conversations about digital media, and joint-media engagement frequency. Participants include children aged 5 to 8, with various gender identities, and parents aged 34 to 47. The table also notes how often families engage in conversations about digital media and participate in media activities together, with responses ranging from "Sometimes" to "Always." The data shows a diversity of experiences and engagement levels across different families.}
    \label{tab:dem}
\end{table*}

\subsection{Participants}
We recruited a total of 20 parent-child dyads to participate in a 75 minute virtual study. We excluded two additional participants because they failed to follow experimenter directions and did not complete all study activities. We required that parents and children be fluent in English, and that children be between the ages of 5--8 (note: we only ran one session with 5-year-old participant). Table \ref{tab:dem} includes a description of participants' demographic backgrounds, whether children had an SEL program at their school or not, how often parents reported having conversations about digital media with their child, and how often parents reported consuming media jointly with their child. We recruited families from a large technology company, and participants completed the study using their own iPads or company loaner devices. Families were compensated with \$12 meal vouchers for parents and \$10 gift cards for children.

\subsection{Study Procedure}
We conducted 75-minute study sessions remotely over video conferencing software. At the beginning of the study, we obtained informed consent before asking parents to download the app onto their devices. The child was given one of two videos to watch in counterbalanced order. After the child finished watching one of the videos, they either completed a generated activity or were assigned to the no activity condition (again, in randomized order to account for ordering effects). After the video or activity, the child was asked to: ``\textit{Please tell me what happened in the video you just watched}.'' 
This process was then repeated for the second video in the other condition. After children completed both tasks, they gave open-ended feedback about the study activities. 

While the child watched the episodes and completed the activity, we asked their parent to move to another room or listen to music through noise-cancelling headphones, so they would not hear what was happening in the show. During this time, the parent completed a survey including parent-child relationship, age, and gender information. We additionally asked about the parent's involvement in their child's social-emotional learning and the nature of their conversations around mediating screen time. Finally, we asked how often the parent and child jointly consumed media based on the Parent Play Questionnaire \cite{ahmadzadeh_parent_2020} to understand when and why parents use devices with their children. 

After the child completed both videos and the activity, the parent viewed the parent interface and provided feedback through a semi-structured interview (Section \ref{sec:study_design}). Questions focused on how the parent interface may or may not promote deeper SEL or content-based discussions, relevance of the technology to them as a parent and the connection they have with their child, and potential use of eaSEL in daily life. We deliberately chose to ask parents for direct feedback rather than observing a parent-child conversation supported by this interface because we felt that observer effects would have too great of an impact on an in-study conversation.

\subsection{Data and Analyses}

Our data included surveys from parents, interaction data, feedback from child participants, and semi-structured interviews with parents. All surveys and interview questions are included in Supplementary Material. Parent background information is shown in the participant table, and we analyzed other data as follows.

\paragraph{Children's Video Retellings}
Prior works indicate that affect labeling (putting emotions into words) and the use of emotion vocabulary are tied to both emotion regulation and development \cite{varatharaj_children_2024, lindquist_language_2021, torre_putting_2018}, so to evaluate children's comprehension of and attunement to SEL topics, we measured the emotion vocabulary in their retellings of the video plot. We collected and manually transcribed recordings of children's re-tellings of the TV episodes they watched, and extracted features related to emotion word usage (general affect words, positive emotion words, and negative emotion words) using the LIWC15 lexicon \cite{pennebaker_linguistic_1999}. 
We then quantitatively compared the presence of affect words and emotion word proportions (number of unique emotion words / unique words in full re-telling) across the \textit{No Activity} and \textit{eaSEL Activity} conditions using a Wilcoxon signed rank test for non-normal data. Note that affect words include terms related to general emotional processes (e.g., "feel", "feeling") in addition to positive (e.g., "kind", "hug") and negative emotion words (e.g., "angry", "fight").

\paragraph{eaSEL Activity Artifacts}
Our app automatically collected information on the generated activity prompts and the child's response to the activities. To understand \textit{how} generated activities stimulated children's social-emotional learning, we qualitatively analyzed artifacts (i.e., drawings) created by the children. To this end, we manually transcribed all the artifacts with recordings, and then identified themes across artifacts using a bottom-up approach.

\paragraph{Children's Open-Ended Feedback}
After children completed both videos, we asked them about their likes and dislikes of the study overall. Their responses were transcribed manually.

\paragraph{Parent Interviews}
We transcribed the semi-structured interview recordings and analyzed those transcripts using reflexive thematic analysis. Our analysis is semantic and realist, using a mixed inductive and deductive approach to the data \cite{braun_using_2006}. The first author led analysis by reading through interview transcripts, taking notes, and synthesizing an initial codebook which was shared with the other authors, who then provided feedback to the initial codebook. From this feedback and continued iteration, the author refined codes; the final codebook is provided in Supplementary Material. From parent interviews, we report themes that emerged from this coding process.

\section{Results}
In the following section, we report mixed-methods results related to each of the two research questions about child-focused activities and parent-oriented information.

\subsection{RQ1: Child's Social-Emotional Learning and Enjoyment}

\begin{figure*}[t!]
    \centering
    \includegraphics[width=.8\textwidth]{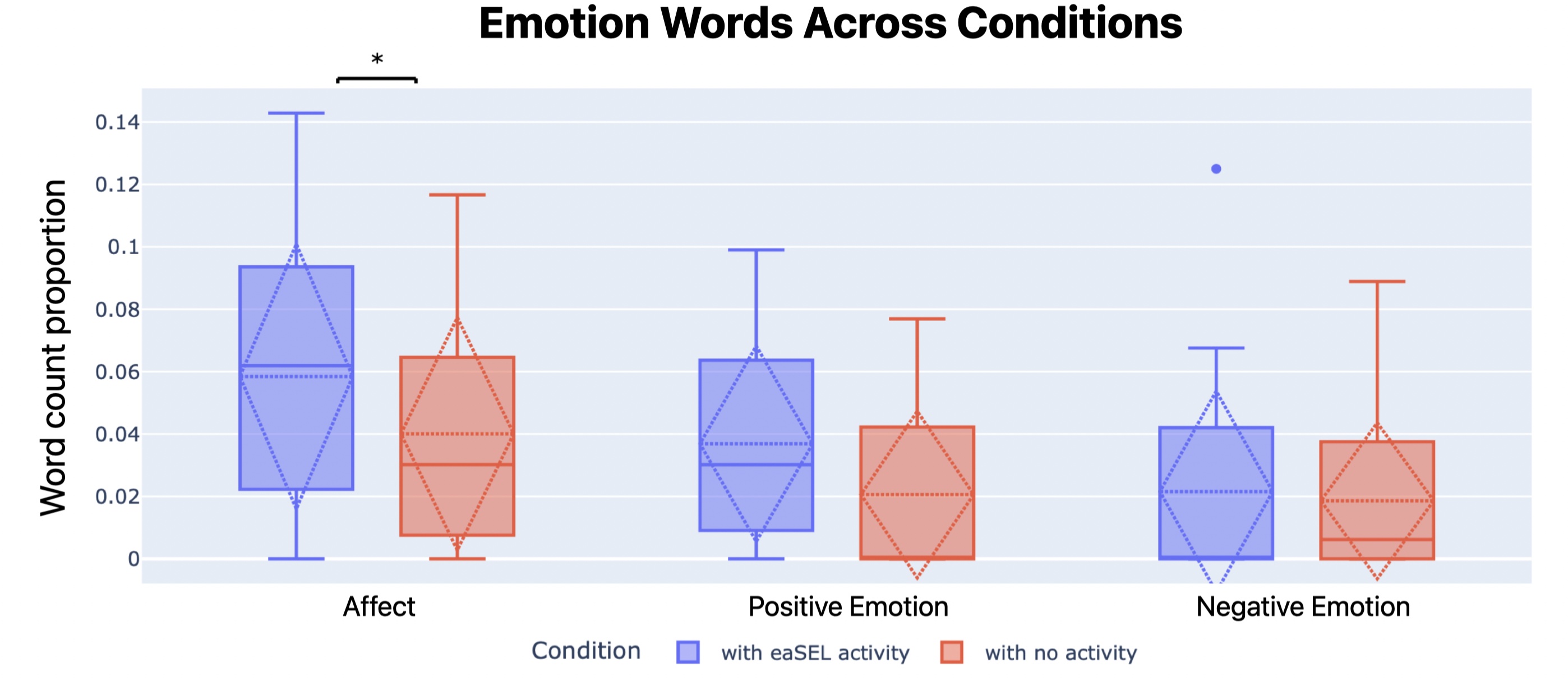}
    \caption{Emotion word counts in children's story retellings across conditions. The eaSEL activity condition resulted in significantly more overall affect words during retelling than the no activity condition (* denotes a difference at p < .05; N=20 children).}
    \Description{The figure shows a box plot comparing the proportion of unique of emotion words used by children out of total unique words in story retellings across two conditions: with and without an SEL activity (eaSEL). The plot is divided into three categories: affect (general affective process/mood related words), positive emotions, and negative emotions. The frequency of affect words is higher in the condition with SEL activity compared to the condition without, as indicated by a significant difference marked with an asterisk.}
    \label{fig:emotion-words}
\end{figure*}
\paragraph{Emotion Words in Children's Video Retellings}
As shown in Figure \ref{fig:emotion-words}, children use proportionally more affect words ($p = 0.02$, Cliff's d = $0.27$/small) in their story retellings after doing eaSEL activities ($M = 0.058$, $SD =  0.042$), compared to just watching the video ($M = 0.040$, $SD = 0.037$). For example, C13 used the affect words ``kind'' and ``hugged'' in their retelling: ``\textit{I think it was being \textbf{kind} to friends, and then they \textbf{hugged} and now the girl’s it}''. We do not find any difference in the frequency of positive (e.g., ``\textit{I \textbf{love} my stuffy and started \textbf{kissing} it}'' (C21)) or negative (e.g., ``\textit{People got, like, I think \textbf{angry} and they start, like, not following the rules}'' (C06)) emotion words across conditions.

\begin{figure*}[t!]
    \centering
    \includegraphics[width=.75\textwidth]{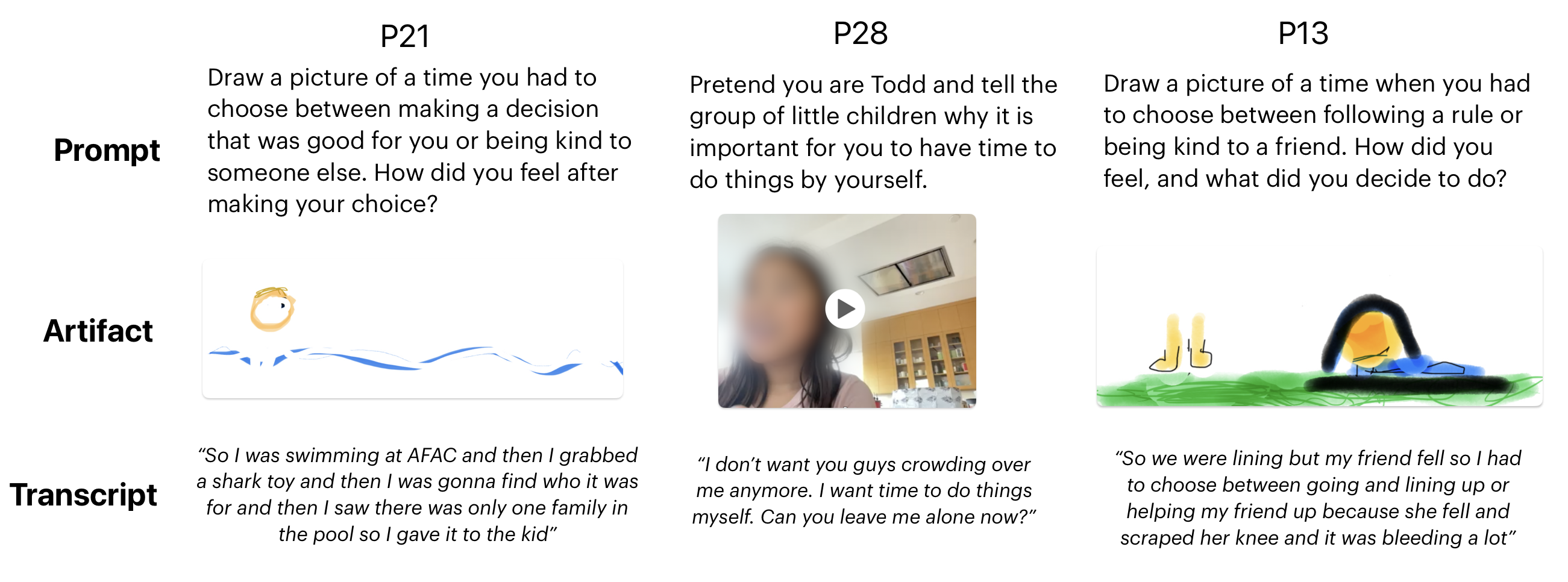}
    \caption{Examples of artifacts produced by children, including transcripts.}
    \label{example_artifacts}
    \Description{The figure presents examples of artifacts produced by children in response to prompts, along with their corresponding transcripts. Three prompts are shown: one asks a child to draw a picture about making a difficult decision, another asks a child to role-play as a character explaining the importance of alone time, and a third asks for a drawing about choosing between following a rule or helping a friend. Each prompt is paired with a child's artifact—either a drawing or a video—and a brief transcript of the child explaining their creation or decision. The artifacts and transcripts illustrate the children's understanding and responses to the prompts.
}
\end{figure*}

\paragraph{Analysis of Child Artifacts}
Examples of artifacts created by children are shown in Figure \ref{example_artifacts}. Overall, children understood the activities without needing extra support from the researcher, and only four children gave short, few word answers (e.g., ``\textit{scared face}'' (C08) or ``\textit{by saying sorry}'' (C07)). These shorter responses still addressed the presented activity prompts, but demonstrate shallower engagement with the SEL lessons those activities were meant to invoke. 

The remaining 16 children gave more thorough answers. In these activities, children demonstrated their social emotional knowledge primarily through the modes of personal experience sharing or role playing. For example, when prompted to reflect on a time he had to decide between helping himself and helping someone else, C13  shared a story about reuniting a found toy with its owners, \textit{“So I was swimming at AFAC and then I grabbed a shark toy and then I was gonna find who it was for and then I saw there was only one family in the pool so I gave it to the kid.”} 

C28 instead engaged in a role-playing activity where she took on the role of a character in the show that needed some personal space. She responded with: \textit{“I don’t want you guys crowding over me anymore. I want time to do things myself. Can you leave me alone now?”} Regardless of whether children engaged in personal storytelling or role-playing, their activity responses demonstrate components of reflection (e.g., examining feelings or taking on other perspectives).

 \paragraph{Children's Qualitative Experiences}

When asked about their experience with the activities, children commented on enjoying all modalities of the activities (drawing, recording, and taking videos), and also offered suggestions for more activities and features. For example, C08 said, ``\textit{I liked answering, like, the, the activities. I like those,}'' while C24 shared, ``\textit{I liked that I got to act things out, and I didn’t like...I liked everything.}'' While C06 also liked the drawing activity (\textit{``I really like drawing''}), they also had suggestions for adding to the activities: \textit{``Maybe [...] more activities. Drawing part, maybe like, we could like add some effects to it and some sound effects.''} This feedback offers initial evidence that eaSEL provides an enjoyable way for children to engage more deeply with SEL content in videos. 

\subsection{RQ2: Parent-Child Interaction via eaSEL}
Two primary themes emerge from our parent interviews: (1) enhancing children's active reflection practice to drive real-world learning and (2) scaffolding deeper parent-child conversations.

\subsubsection{Theme 1: Active Reflection Practice to Drive Real-World Learning}

Parents find that eaSEL encourages \textit{active} engagement with content that their child may otherwise engage with more passively, explicitly mentioning that the tool promotes meaningful reflection. P33 shared that eaSEL allows kids to \textit{“step back and think about what is this trying to tell, what is the story about, [and] what skill sets [the story is] trying to nurture.''} They followed with, “\textit{I like that because...we just mindlessly watch a lot of shows without a goal. So [this tool]'s also having something constructive, something with a purpose and being able to see it and track it and control it in a way.”} By interleaving video content with activities, children are encouraged to pause and think in a way that directly promotes reflection.

Furthermore, most parents felt optimistic about how eaSEL-guided reflection could support a regular practice of engagement with social-emotional topics. P18 appreciated how the presence of games or activities could encourage a child to utilize these crucial skills more regularly: \textit{“It would really help them. Cause unless you practice it, you don't use it every day, even though you know it, you know, like practice makes you better.''} Similarly, when asked about ways in which the tool could support SEL, P27 shared ``\textit{I think that everything takes practice and the more you engage and talk about emotions and situations and how you would, deal with certain scenarios that may not have happened yet but are definitely likely to happen, is a step in the right direction...it's all about practice.}'' Only by practicing skills do people build the confidence to use them, and parents appreciated how the tool could facilitate social-emotional learning through regular reflection. Ultimately, parents found that the opportunities for practice could lead to better long-term outcomes when watching videos, such as ``\textit{being more thoughtful when they consume}'' (P08). Overall, parents expressed how using eaSEL could encourage regular practice of active reflection, which could drive social-emotional learning.

Finally, parents expressed how this active practice may translate to real-life skill development outside of the media-consumption context because eaSEL serves as a bridge between the abstract parts of the show and the ``\textit{learnings of the real world}'' (P39). Specifically, by drawing connections between the show and a child's life, or by asking the child to put themselves in the shoes of a character, children can more easily connect the dots between the content, the lessons, and themselves. For example, P10 shared that the tool ``\textit{really tie[s] the message because watching a show and not really thinking about it or talking to someone about it may not really be a connection to, like, the moral of the story. So I think this is a really good way to kind of like reinforce that learning.}'' These observations connect to our goal of providing content-grounded activities, as parents highlight how eaSEL turns children's media consumption from passive watching into an active learning process grounded in diverse modalities and real-world scenarios. 

Overall, eaSEL supports reflection on SEL content within videos. Parents appreciate this active engagement, the opportunity to practice social-emotional reflection, and the potential for real-world learning benefits that might come by drawing connections between the content and the child's lived experiences.

\subsubsection{Theme 2: Scaffolding Deeper Conversations Between Parents and Children}

The parent interviews also showed that by providing context beyond summaries, the tool can help scaffold potential conversational touchpoints and support parental learning and teaching. When asked about their first impressions of the parent interface, some parents appreciated the summary of the video the child had watched, such as``\textit{...I feel like I caught up with what he watched, so I have a pretty good understanding}'' (P27). The summary provides the content overview to support conversation without necessitating co-viewing, which gives parents an opening to conversation. However, some parents commented that for long-term use, extensive summary information about the show---a common approach in extant parent dashboards---could become overwhelming. For example, P11 mentioned ``\textit{I think there's a lot of text, takes a long time to read, so I think using less text, more bullet points, more shorter formed kind of statements would be easier to digest.}'' 

In contrast, parents shared that they particularly enjoyed seeing the artifact their child had created. For example, P38 mentioned that the ``\textit{most engaging [part] to me [was] trying to see  what she drew}'', and P08 echoed, ``\textit{I'm drawn to what they came up with in the moment. So the drawing and the audio recording is, I would argue, most engaging...anytime they draw something or record something like I'm always into that.}'' This interest in the child's creative artifacts could drive parent engagement, but also give parents more insight to adapt to their child's specific SEL needs: ``\textit{For me is to find out more on what's going on with [my child] at school and outside of this house...The teacher will tell me about how she's doing her math or language or geography... I don't think the teachers give me a social emotional learning report like how is [my child's] behaviors}'' (P08). Parents recognize that as children grow, they have many experiences outside the home, and seeing how children might think about interacting with someone who is not a family member can also inform parent-child conversations around SEL.

While parents wanted to deepen these SEL-based discussions with their children, parents had varying degrees of SEL knowledge themselves. In some cases, parents mentioned that the tool's social-emotional skill definitions helped them learn as well and in doing so they felt more equipped to discuss this information with their child. For example, parents shared that the tool helped them ``\textit{have all the, the technical terms and definitions,}" (P24) that they otherwise felt less informed about. P13 shared, ``\textit{I don't think I'm any expert in social and emotional learning, so anything that I can do to be softer around the edges about all these things is definitely something that's beneficial for him and, and probably beneficial to me as well.}'' By being more informed about SEL, parents felt more equipped to support their child, and that knowledge then supports deeper parent-child conversations that relate to lessons rather than plot, and make it easier for the parent to ``\textit{reinforce or echo this lesson}'' (P14). By including summary information on key social-emotional skills, parents can tailor their conversations toward these lessons.

Even beyond directly teaching their children social-emotional skills, parents found the system helpful for ``\textit{highlight[ing] the touch points}'' (P06) more broadly, which could help to deepen their relationship with their child. For example, P38 shared that the parent conversation starter ``\textit{gives me some perspective really. I can start having deeper conversation[s] with my child in situation[s] like this, asking her different roles and how how she would feel about that, which I've never thought of before.}'' 
In particular, many parents noted how they could embed their own experiences into such conversations, ``\textit{There are certain situations where I bring my experiences into it and I talk about what happened to me, but if this is prompting me, I'm more likely to do that}'' (P27). P33 similarly shared that eaSEL``\textit{would allow me to double down on the concept with my own behavior and example.}'' 
Parents note the importance of integrating their own experiences, as one participant shared ``\textit{depending on your upbringing, your cultural background, sort of your own norms and values, you may notice different nuances in the behavior, in the body language that are relevant for you and that maybe are relevant to your child, but that for a general audience may not be appreciated enough to end up in the [generated] summary}'' (P21). These reflections illustrate how the tool provides parents with the tools and prompts to engage in SEL discussions, but that the generalized summaries from LLMs may need to be further personalized with the parent's own values and experiences. As such, our findings align with our goals of supporting diverse parent involvement, promoting parent-child bonding, and bringing content and conversation into the real world. 

\section{Discussion}
As children increasingly consume digital media independently \cite{kabali_exposure_2015, movie__tv_reviews_for_parents_common_nodate}, parents look for technologies that can benefit the child socially and emotionally \cite{neumann_young_2015}. From prior work, we identified a few key challenges in existing digital mediation and SEL solutions. Specifically, SEL curricula highlights the importance of contextually-relevant at-home SEL practice \cite{slovak_teaching_2015}, but HCI efforts in this space are often divorced from current activities (e.g., require separate engagement with a chatbot) and therefore lack contextual relevance \cite{seo_chacha_2024} or rely on joint-media engagement, which is not always realistic for parents and families \cite{hiniker_screen_2016, zhang_storybuddy_2022}. This work fills these gaps by (1) supporting active, independent SEL engagement during video-watching by integrating reflection activities with the storyline of children's shows and (2) scaffolding parental engagement post-hoc through conversation-starters. 

Our findings show that eaSEL promotes social-emotional reflection through AI-generated children's activities and provides scaffolding that parents perceive could deepen family conversations. The first part of these findings are supported by our quantitative analysis, where children demonstrate higher emotion word usage in story retellings after completing eaSEL activities (supporting [\textbf{H1}]). Our qualitative exploration of child-produced artifacts reveal cases of effective social-emotional demonstration, and child participants express enjoyment of the activities. Regarding findings related to our second research question, we reveal parents' positive evaluations of the system in supporting children's learning through active practice and in providing appropriate scaffolding that could stimulate deeper family connection. In the following subsections, we discuss broader challenges and insights revealed by our work.

\subsection{Meeting Families in Independent Use Contexts}
A key driving principle in this work was supporting learning and reflection in the context children most often consume media---independently watching videos. While prior work often focuses on joint-media engagement solutions \cite{zhang_storybuddy_2022, smith_contextq_2024}, we recognize that co-viewing is not always realistic for families \cite{movie__tv_reviews_for_parents_common_nodate}. Instead, we demonstrate that systems can support meaningful and independent active learning for child users by directly integrating with video content. Through our user study, we see that after completing post-video reflection activities, children demonstrated content reflection and better understanding of social-emotional situations from content they watched. Parents similarly expressed positive sentiments about using eaSEL to support their child's independent active reflection practice to consume media more mindfully. However, as we note in Section \ref{systemlimitations}, new challenges related to content mediation arise from this independent approach, as certain unsafe or inappropriate stories should not have corresponding social-emotional lessons. This tension could be explored in future works related to child-focused guardrails or content filtering algorithms \cite{saldias_designing_2024}, or by supporting human in the loop moderation of the content and activities that are presented to children. This work illustrates how AI-driven interventions can facilitate reflection and scaffold learning, but also highlight the need for more robust model safety and reliability to move this degree of AI engagement into scope.

\subsection{Supporting At-Home SEL with Parental Involvement}
Our second key driving principle was promoting SEL curricula at home, with a heavy emphasis on parental involvement post-hoc. While SEL is primarily formally taught at school, it is important to involve parents in supporting learning in the family context as well \cite{belsky_determinants_1984, li_how_2023, yucel_siblings_2015, dunn_family_1991}. As we identified in our user study, parents have varying degrees of knowledge around SEL. Prior work focused on how workshops might support parent learning \cite{darling_social_2019, mccormick_effects_2016, slovak_scaffolding_2016}, but low turnout with such workshops is a challenge that could be supplemented by incorporating technology-driven SEL at home.

We show that eaSEL helped parents learn as well, making them feel that they would be more equipped to have discussions about SEL with their child. We accomplish this goal by providing parents with explicit SEL information so they have the context, confidence, and repertoire to engage on these topics. Then by showing their child's SEL activities and providing conversation starters grounded in personal experience, we give them information they feel they could use to begin SEL-grounded discussions that do not rely on a deep knowledge of the video content itself and maintain relevance to the relationship between the parent and child. 

When introducing SEL at home, we found through parent interviews that diverse family values (e.g., cultural differences) should be respected, further emphasizing the important role of incorporating parent's voices in this process. For example, parents shared that language models might normalize or overlook moments in shows that are particularly relevant to their cultural background. Our work reveals the tension between minimizing parent load through system autonomy and contextually-grounded conversation starters, while still directly integrating the parent's preferences in both \textit{what} they want to teach and \textit{how} they want to teach it. As we look to the future, one novel solution might employ the SEL detection task to recommend rich media to parents, who can determine whether the lessons of the show are personalized towards their parenting preferences and family values.

\subsection{Bringing eaSEL Into The Real World}

To make both child activities and the parent interface information practical for family contexts, we also imagine a more integrated and networked future implementation. Specifically, for simplicity during our user study, we presented the child activity and parent interface on the same device. Furthermore, although our activity generation pipeline could in theory generate activities for any video and therefore integrate into video-watching apps children already use, we conducted this research in a standalone app with a curated set of videos. To bring eaSEL in its current form into the real world, we imagine a solution where a child could watch videos and complete activities on whatever device and in whatever app they regularly use, and the parent might receive the relevant information from the interface on their own device (e.g., via push notification, text message, or email).

By presenting both child activities and contextual information to the parent in places where their intended audiences will see them, these two content reflection entry points could act as fail-safes to one another: if a child chooses to skip or skimp on an activity, a parent might notice and choose to more deeply engage with their child around that topic. Likewise, if parents are too busy to converse with their child about the content they watch, the child activities could still help the child to engage in some independent meaningful reflection. In the future, we hope to explore how a fully integrated and networked implementation of eaSEL could be used in-the-wild.

\subsection{Study Limitations and Future Directions}
We opted to conduct a single-session study of eaSEL as an initial step to demonstrate the potential for technology to achieve our two key driving principles of promoting children's reflection in independent use contexts and supporting at-home SEL with parental involvement. We note, however, a few limitations from our study design. First, our study consisted of a relatively small sample with limited diversity, as all parents are in a similar age group, work in the technology industry, often already discuss media consumption with their children, and are mostly aware of SEL curricula at their child's school.
Parents already interested in SEL or digital content mediation might have been more likely to sign up for the study, resulting in some self-selection bias. Future work should explore whether LLM-driven SEL solutions provide benefits to users of more diverse backgrounds.

Second, the short evaluation period meant that although we could evaluate children's activity artifacts, children's emotional language, and parent perceptions of conversation starters, we were not able to study longitudinal learning outcomes or the parent-child conversations themselves. In the future, we would like to run a long-term at-home deployment with a larger, more diverse sample using a more integrated and networked eaSEL implementation, like the one described in the prior section. Such a study would enhance the reliability of our findings and bridge the gap between our current implementation of eaSEL and future versions applied to a wider range of videos. In doing so, we hope to explore the impacts that long-term use of our system may yield, better understand the types of conversations parents and children have when using the system, and how parent and child behavior with eaSEL may change over time.

\section{Conclusion}

In this work, we presented eaSEL, an AI-driven system that (1) presents social-emotional learning (SEL) activities alongside children's videos and (2) provides conversation-starters for parents to deepen discussion around digital media and social-emotional learning themes with their child without necessitating co-viewing. We draw on SEL curricula to introduce and benchmark large-language model tasks for (a) identifying SEL moments in children's videos and (b) generating parent and child activities. Then, we conducted a mixed-methods study with $N=20$ parent-child dyads demonstrating the system's effectiveness in enhancing children's SEL reflection and showing how parents believe that contextual information and conversation starters could deepen future conversations around digital media and social-emotional themes. We found that children used more emotion words in story retellings after completing eaSEL activities compared to watching media alone and child-produced artifacts demonstrate children engaged with social-emotional concepts. Parents find that eaSEL is helpful for promoting children's active reflection and could scaffold deeper dialogue within families. Overall, our work paves future directions in exploring how AI can support children's social-emotional development and strengthen family connections over media consumption in the digital age. 

\begin{acks}
We would like to thank our participants for their invaluable contribution to this study. We would also like to thank R. Benjamin Shapiro, Katie Van Sluys, Dianna Yee, and Mary Beth Kery for providing feedback to this project.
\end{acks}

\bibliographystyle{ACM-Reference-Format}
\bibliography{sample-base}

\newpage
\appendix
\section{Prompts} \label{prompts}
\subsection{SEL-Moment Detection Prompt}
\begin{MyVerbatim}
We will provide a transcript of a kid’s video. First, think about the major plot points of the video. 
Then, determine whether this transcript contains a moment related to social-emotional learning (either positive or negative) for each of the major plot points. 
If you think that multiple skills could apply for a single plot point, pick the one that is the most suitable.

In particular, identify if the transcript contains any moment where a character demonstrates the skill: [SKILL] or fails to demonstrate the skill: [LACK_OF_SKILL].
We define this skill as [SKILL_DEFINITION]

A positive example of this skill is: [POSITIVE_EXAMPLE]. A negative example of this skill is: [NEGATIVE_EXAMPLE].

If the transcript contains a positive or negative example of this skill for one of the major plot points, reply with a 1. Remember that each major plot point should only have at most 1 social-emotional learning moment. Note that some major plot points won't demonstrate any social emotional skill. Then, after a comma, provide a short 1-2 sentence explanation of why the transcript contains a moment related to social-emotional learning and why that moment is a central plot point. Otherwise reply with a single 0. Let's think step by step.

Tips while annotating:
* If you’re uncertain or on the fence, review the definition and notes before deciding
* Something is a “main” event if:
    * It would appear in a 2-3 sentence summary of the episode 
    * If the moment were dropped would the story have changed?
    * Would this plot point make 2 separate questions? (2 separate SEL skills)?
* Skills should typically be made explicit by the transcript

Now read the following transcript, and think about the major plot points and social-emotional learning moments related to the major plot points: 
[TRANSCRIPT]

Skill (0 - not present or 1 - present and is part of the central plot), explanation (if 1 - present). Separate the rating and explanation with a COMMA:

\end{MyVerbatim}

\subsection{Child Activity Generation Prompt}
\begin{MyVerbatim}
SEL_GENERATION_PROMPT_DRAWING = """
You will be given a transcript from a children’s video that contains a social-emotional learning moment related to a skill (the skill will be described). You will then provide a short summary of the social-emotional event that occurred and generate an activity that asks the child to draw a picture of a time in their life where a similar event happened as in the media they consumed. The activity should encourage the child to practice/reflect on the social-emotional skill. Make sure to include examples for the child when applicable. Ensure that only ONE question is asked, and do NOT generate anything other than the question alone. Let’s think step by step.

Example input:
Transcript: “Frog saw Toad’s ice cream looked more delicious, so he ran over and stole Toad’s ice cream.”
Social-emotional skill: consideration for others

Example response:
“In the video you just watched, Frog took Toad’s ice cream. Draw a picture of a time someone took something from you and how that made you feel. For example, if someone knocked down a sand castle you built, took your place in line at the cafeteria, or stole a pencil from your pencil box.”
"""

SEL_GENERATION_PROMPT_IMAGINE = """
You will be given a transcript from a children’s video that contains a social-emotional learning moment related to a skill (the skill will be described). You will then provide a short summary of the social-emotional event that occurred and generate an activity that asks the child to imagine or predict different story outcomes based on what happened in the transcript. The activity should encourage the child to practice/reflect on the social-emotional skill. Ensure that only ONE question is asked, and do NOT generate anything other than the question alone. Let’s think step by step.

Example input:
Transcript: “Frog saw Toad’s ice cream looked more delicious, so he ran over and stole Toad’s ice cream.”
Social-emotional skill: consideration for others

Example response:
"In the video you just watched, Frog took Toad's ice cream. Imagine the ice cream melted before Frog could give it back. How could Frog make it up to Toad?"
"""

SEL_GENERATION_PROMPT_STORY = """
You will be given a transcript from a children’s video that contains a social-emotional learning moment related to a skill (the skill will be described). You will then provide a short summary of the social-emotional event that occurred and generate an activity that asks the child to tell a personal story of a time in their life where a similar event happened. The activity should encourage the child to practice/reflect on the social-emotional skill. Make sure to include examples for the child when applicable. Ensure that only ONE question is asked, and do NOT generate anything other than the question alone. Let’s think step by step.

Example input:
Transcript: “Frog saw Toad’s ice cream looked more delicious, so he ran over and stole Toad’s ice cream.”
Social-emotional skill: consideration for others

Example response:
“In the video you just watched, Frog took Toad’s ice cream. Tell me a time where someone took something from you and how that made you feel. For example, maybe someone was hogging the seat to the swing, or a friend or sibling took your toy.”
"""

SEL_GENERATION_PROMPT_ACT = """
You will be given a transcript from a children’s video that contains a social-emotional learning moment related to a skill (the skill will be described). You will then provide a short summary of the social-emotional event that occurred and generate an activity that asks the child to role play/act out how they would respond to a specific scenario in the transcript. The activity should encourage the child to practice/reflect on the social-emotional skill. Ensure that only ONE question is asked, and do NOT generate anything other than the question alone. Let’s think step by step.

Example input:
Transcript: “Frog saw Toad’s ice cream looked more delicious, so he ran over and stole Toad’s ice cream.”
Social-emotional skill: consideration for others

Example response:
“In the video you just watched, Frog took Toad’s ice cream. Act out how you would respond to Frog if you were Toad and your ice cream just got stolen.”
"""

SEL_GENERATION_PROMPT_SUFFIX = """

Criteria for a good generated activity:
- age appropriate
    - The activity uses language appropriate for a 5-8 year old (not too many 3-4 syllable words, simple sentence structures, and topics children are familiar with). For example, instead of "perseverance" you can say "never giving up"
- structure:
    - The prompt contains a short summary of the social-emotional event BEFORE introducing the activity
    - The activity is not a composite of multiple questions/activities
    - Refer to the character names directly instead of just saying "the character"
- relevance:
    - The activity relates to the transcript
    - The activity relates to a social-emotional skill in the transcript
- promotes social-emotional learning
    - The activity encourages the child to reflect on social interactions or emotions of themselves or others OR to put themselves into the shoes of others, including characters in the story

Now generate an activity for each of the following skills + activities. ONLY include the generated activity, with no other additional information (like "Activity:", type of activity, etc):
Transcript: [TRANSCRIPT]
Social-emotional skills: This transcript contains the skill [SKILL_DESCRIPTION], because [SKILL_EXPLANATION]
Activity:
"""
\end{MyVerbatim}

\subsection{Parent Activity Generation Prompt}
\begin{MyVerbatim}
You will be given a transcript from a children’s video that contains a social-emotional learning moment, and what social-emotional skills are demonstrated or not demonstrated in the segment. Please generate a conversation starter to give to a parent, so that they can have a discussion with their child about the social emotional topic present in the transcript. The open question/conversation starter should involve telling a personal story/experience related to the social-emotional topic. The conversation starter must include an open question or conversation starter related to the parent's personal experiences. Generate one conversation starter based on the social emotional topic and the transcript. Make sure only one question is asked, and do not include nested questions. Let’s think step by step.

Examples:
Transcript: “Frog saw Toad’s ice cream looked more delicious, so he ran over and stole Toad’s ice cream.”
Social-emotional skill: social skills
Social-emotional skill explanation: This transcript demonstrates a moment of social-emotional learning related to "social skills" because Frog does something that is not kind to Toad (stealing his ice cream).
Parent activity prompt: Before bed, tell your child a story about a time where someone took something that belonged to you.

    Examples: For example, maybe a co-worker received a promotion you thought you were going to get, or a family member ate your food in the fridge.

Transcript: "Stillwater tells Carl to take a deep breath after he goes on a rant about how angry he is"
Social-emotional skill: regulating negative emotions
Social-emotional skill explanation: In the transcript, Stillwater teaches Carl how to regulate his anger by taking deep breaths.
Parent activity prompt: When you get the chance, tell your child about a time when you had to hold your emotions in and how you resolved it.

    Examples: For example, maybe you were nervous before a presentation and took deep breaths or you wanted to laugh at a joke in a meeting but held it in.

Criteria for a good parent activity prompt:
- make sure the prompt asks the parent to discuss with the child, rather than just asking the question directly to the parent (ie. instead of "Talk about a time when...", you should respond with "Tell your child about a time when...")
- provide a few examples to stimulate memories, and make sure to format the examples as Examples:...
- make sure that the social-emotional skill is made explicit to the parent

Now generate a parent activity given the following:
Transcript: [TRANSCRIPT]
Social-emotional skill: This transcript contains the skill [SKILL_DESCRIPTION]
Social-emotional skill explanation:[SKILL_EXPLANATION]
Parent activity prompt:
\end{MyVerbatim}

\section{Full SEL Skills}\label{full-sel}

\onecolumn
\footnotesize
\begin{longtable}{| p{.1\textwidth} | p{.05\textwidth} | p{.1\textwidth} | p{.25\textwidth} |p{.2\textwidth} |p{.2\textwidth} |} 
 \hline Area & Skill ID & Skill & Definition & Positive Example & Negative Example 
 \\ \hline Self-awareness & A1 &  Identifying one's own feelings 
& 
\begin{itemize}
    \item The ability to recognize, understand, and label emotions in oneself.
    \item Understand the definition of feeling words using self as an example
    \item Identify feelings based on face and body cues, and context
    \item Monitor intensity of feelings
    \item Identify situations that you anticipate may trigger certain feelings in the
    \item future
    \item Understand that you can have multiple feelings at once
    \item Within one character
\end{itemize}
 & 
Max sees Joe walk away, and his chest pangs. 
"I must be sad that Joe is leaving," Max 
thinks.
 & 
Max sees Joe walk away, and his chest pangs. Max feels 
irritated and snaps at a nearby dragonfly, "Why do you 
have to buzz around so much?!" Instead of recognizing 
his sadness, Max misinterprets his emotions as 
annoyance and takes it out on others.
 
\\ \hline Self-awareness & A2 & 
Having accurate 
self-knowledge
 & 
 \begin{itemize}

    \item Developing and maintaining a coherent understanding and sense of oneself
    \item over time, including
    \item personality traits, interests, preferences, strengths, and weaknesses
    \item Identifies and understands personality/character traits
    \item Recognizes and understands one's own strengths and weaknesses (including
    \item personal biases)
    \item Identifies and understands one's interests and preferences (i.e., likes, dislikes,
    \item desires, preferred learning style, etc.)
    \item Explores, develops, and maintains a coherent sense of self and roles over
    \item time and across different contexts and social identities (including racial,
    \item ethnic, gender, sexual, religious, etc.); tries to remain authentic to oneself
    \item Is honest about what one does and does not know
    \item Scoped to social and emotional topics
    \item Within one character
\end{itemize}
 & 
Max enjoys drawing, but often felt his skills 
weren't good enough. Joe compliments Max's 
sketch. Max reflects, and thinks, "I do have a 
good eye for detail and a steady hand."
 & 
Max enjoys drawing, but often felt his skills weren't 
good enough. Joe compliments Max's sketch. "He's 
probably just being nice, and I actually suck at drawing," 
Max thinks.
 
\\ \hline Self-management & M1 & 
Regulating negative emotions including impulse, stress, aggression, and pessimism
 & 
 \begin{itemize}
  
    \item Ability to use effortful control strategies to moderate one's emotional
    \item reactivity (e.g., to cope with aversive
    \item feelings) and/or automatic behavioral responses
    \item Self regulation of negative emotions
    \item Have a chance to have a reaction and respond in a way against a negative
    \item reaction
    \item Within one character
    \end{itemize}
 & 
Max is angry because a classmate was mean 
to him. Max sits by himself and takes a few 
deep breaths to calm himself down before 
going out to eat with Joe.
 & 
Max is angry because a classmate was mean to him. 
When Joe asks what's wrong, Max snaps at him angrily.
 
\\ \hline Self-management & M2 & 
Displaying grit, 
determination or 
perseverance
 & \begin{itemize}

    \item Shows motivation, determination, or passion to complete tasks and goals
    \item Having courage, resolve or strength of character
    \item Use positive self-talk to provide encouragement when working
    \item toward a goal
    \item Takes action and shows initiative to accomplish an established goal
    \item Persist when encountering something challenging
    \item Oneself or as a group
    \end{itemize}
 & 
Max is almost at the finish line. His chest 
hurts from running, but he thinks to himself, 
"I can make it!" Max finishes in record- 
breaking time.
 & 
Max is almost at the finish line. His chest hurts from 
running, so he thinks "I can't do it..." Max gives up and 
lies down on the ground, unwilling to move.
 
\\ \hline Social awareness & S1 & 
Identifying other 
people's feelings
 & 
 \begin{itemize}

    \item The ability to recognize, understand, and label emotions in others.
    \item Understand the definition of feeling words using others as examples
    \item Identify other's feelings based on face and body cues, behaviors,
    \item and context
    \item Understand that other people can have multiple feelings at once
    \item Towards another
    \end{itemize}
 & 
Max sees Joe's face turn red when the teacher 
calls on him. "He must be embarassed," Max 
thinks.
 & 
Max sees Joe's face turn red when the teacher calls on 
him. Max thinks, "Joe must be angry with the teacher." 
Later, Max avoids talking to Joe, thinking he's in a bad 
mood, when in reality, Joe was just embarrassed.
 
\\ \hline Social awareness & S2 & 
Perspective taking/ 
empathy
 & \begin{itemize}

    \item The ability to understand another person's emotional state and point of view.
    \item This includes identifying, acknowledging, and acting on the experiences,
    \item feelings, and viewpoints of others, whether by placing oneself in another's
    \item situation or through the vicarious experiencing of another's emotions.
    \item Recognize that people can have different feelings in response to the same
    \item situation
    \item Describe somebody's point of view (i.e., thoughts) in a situation and/or
    \item consider a situation from different points of view
    \item Predict somebody else's feelings or behaviors based on their point of view
    \item Recognize that people's feelings can change
    \item Action motivated by an understanding of how someone feels
    \item Lack of empathy is knowing how someone feels but still acting in an
    \item insensitive way
    \item Towards another
    \item Consider or predict behavior based on putting yourself in someone else's
    \item shoes
    \item Explicit expression of emotion from a character and another character
    \item acting on understanding of that emotion
    \end{itemize}
 & 
Max notices that Joe never has enough to eat 
during lunch. Max decides to share his lunch 
with Joe everyday. 
Max sees Joe carrying a lot of books. He 
must be in pain carrying those books. "Let 
me help out with that," he says to Joe, 
smiling sympathetically.
 & 
Max notices that Joe never has enough to eat during 
lunch. He rolls his eyes and thinks "who cares?"
 
\\  \hline Social awareness & S3 & 
respecting 
diversity and 
different 
viewpoints
 & 
 \begin{itemize}

    \item Believes it is important to be tolerant and accepting of differences in others;
    \item celebrates/appreciates
    \item diversity
    \item Recognizing the importance of diversity and valuing differences
    \item Scoped to identity/sense of self
    \item Towards another
    \end{itemize}
 & 
Max sees that the other students make fun of 
Joe, because he has bigger spots than 
everyone. Max walks up to Joe and says, 
"hey, I think your spots make you look really 
unique! They're super cool!"
 & 
Max sees that the other students make fun of Joe, 
because he has bigger spots than everyone. Max joins in on making fun of Joe, and says he's uglier than everyone because he's different.
 \\ \hline Relationship skills & R1 & 
standing up for 
oneself
 & 
\begin{itemize}
    \item Is assertive in expressing one's own needs without being aggressive.
    \item make verbal statements that respectfully express a feeling, want, or need
    \item respectful/assertive body language and tone
    \item Stands one's ground when in the face of peer pressure
    \item Within one self
\end{itemize}
 & 
Max is being made fun of by his classmates. 
He stands up, faces them and says, "You guys 
are hurting my feelings. Please stop making 
fun of me."
 & 
Max is being made fun of by his classmates, but he sits 
quietly and takes it, even though his feelings are hurt
 
 \\ \hline Relationship skills & R2 &  
demonstrate social 
skills such as 
helping, giving 
compliments, and 
apologizing
 &  
\begin{itemize}
    \item Ability to organize and navigate social relationships, including the ability to
    \item interact effectively with others
    \item and develop positive relationships. Includes listening, communication,
    \item cooperation, helping, andcommunity building.
    \item Ability to generate and act on effective strategies/solutions to deal with
    \item challenging interpersonal situations.
    \item Example indicators of social skills:
    \begin{itemize}
        
    \item initiating interactions with peers
    \item sharing
    \item turn taking
    \item asking
    \item helping
    \item giving compliments
    \item listening
    \item saying kind words
    \item ask for permission
    \item agreeing
    \item compromising
    \item suggesting an idea
    \item showing interest (verbal + nonverbal)
    \item apologizing
    \item being polite
     
    \end{itemize}
    \item Towards others
    
\end{itemize}
 &  
Max sees Joe wear a new outfit to school. 
"Your outfit looks great!" he says 
Max sees Joe carrying a lot of books. "Hey 
do you need help?" he asks Joe.
 &  
Max sees Joe wear a new outfit to school. "That's weird, 
you're wearing something different today," he says.
 
\\ \hline  
Responsible 
Decision Making
 & D1 &  
Make decisions 
based on moral/ 
ethical standards
 &  
\begin{itemize}
    \item The ability to make constructive, respectful, and healthy
    \item choices based on commonly accepted ethical standards and social norms.
    \item Balances one's needs/desires with personal responsibility, social norms,
    \item safety, and/or ethical standards when making decisions
    \item Within one character
\end{itemize}
 &  
Max is about to steal Joe's lunch, but then he 
thinks "Stealing is wrong."
 &  
Max is about to steal Joe's lunch, but then he thinks 
"Who cares if I steal? Joe can just buy another lunch."
 \\ \hline

\caption{Full list of SEL skills we support in our system}
\label{SEL_skills}
\end{longtable}

\section{Annotation and Evaluation Templates} \label{annotation}

\subsection{SEL Moment Detection Annotation}
For the following questions, (1) rate whether there is any moment central to the plot of the story that is relevant to the social-emotional skill and (2) provide a 1-2 sentence explanation of what the central plot point is and why there is an SEL moment.

\begin{itemize}
    \item Given the same main event/plot point → choose the most representative SEL skill (the explanation only needs to include the major plot point and how it demonstrates the SEL skill)
    \item If you’re uncertain or on the fence, review the definition and notes before deciding.
    \item Something is a “main” event if...:
    \begin{itemize}
        \item It would appear in a 2-3 sentence summary of the episode
        \item If the moment were dropped would the story have changed?
        \item Would this plot point make 2 separate questions? (2 separate SEL skills)?
    \end{itemize}
    \item Assume villainous acts are just part of the story.
\end{itemize}

Note that there may be transcripts that have none of the skills described below. Please respond carefully depending on the transcript.

Transcript: [TRANSCRIPT]

For each of the following skills (providing skill descriptions and examples from the SEL skill list) we asked the following:

Rate how present the social emotional moment is in the transcript
\begin{itemize}
    \item Not present
    \item Present, and is related to central theme of the story
\end{itemize}

Provide a 1-2 sentence explanation of the major plot point and why it's related to the social-emotional skill.

\subsection{Parent and Child Activity Evaluation}

In this task, you will read activity prompts that are generated for a child to do after watching a TV transcript AND prompts that are generated for a parent to have conversation with their child. The child-focused activities are meant to help children develop social-emotional skills. The parent-focused prompts are meant to help parents start conversations with their children. You will read the transcript, then rate a few different child and parent activities. 

You will read activity prompts given to children after watching a TV episode. These prompts are intended to help children engage in social-emotional learning reflection based on what they watched in the show. Make sure to treat tasks as independent (if you've done this task before, do not base your answers off of previous prompts you have read)

\noindent\textbf{Prompt for the child after watching the show: } [CHILD\_ACTVITY]

\noindent This question is relevant to the social emotional learning moment: 

\noindent[GPT\_SKILL\_EXPLANATION]. In particular, rate your agreement to the statement: the prompt for the parent mentions/refers to this moment in the show
\begin{itemize}
    \item Strongly Agree
    \item Somewhat Agree
    \item Neutral
    \item Somewhat Disagree
    \item Strongly Disagree
\end{itemize}

The question asks the child to [CHILD\_ACTIVITY\_TYPE]
\begin{itemize}
    \item Yes
    \item No
\end{itemize}

\noindent The activity can be done by just the child. In particular, to do the activity, the child doesn't need to grab a friend or parent.

Examples:
\begin{itemize}
    \item  (NO to can be done by child alone): Find a friend or parent nearby and act out how you would respond to them if they lost something.
    \item The following response is alright (YES can be done by child alone): Can you tell me about a time when you noticed someone else was scared or upset, and what you did to help them? For example, maybe you saw a friend fall down at the playground and you helped them up or told a teacher.
    \item This example is fine because it's just talking about a friend/teacher -- the activity doesn't require the child to grab another person
\end{itemize}
\begin{itemize}
    \item Yes
    \item No
\end{itemize}

\noindent[SKILL\_DESCRIPTION], [SKILL\_DEFINITION], [SKILL\_EXAMPLES], 
The question is related to the social emotional skill explained.
\begin{itemize}
    \item Strongly Agree
    \item Somewhat Agree
    \item Neutral
    \item Somewhat Disagree
    \item Strongly Disagree
\end{itemize}


\noindent The question promotes reflection in that it...
\begin{itemize}
    \item relates to a personal experience
    \item provides the child an opportunity to view different perspectives
    \item involves the child acknowledging or examining feelings
    \item can provide the child with a basis for change
    \item can allow the child to consider alternative actions
    \item None of the above
\end{itemize}

\noindent The activity is a yes/no question, or can be answered in one word. (For example, "Can you tell me about a time when you were angry?" can be answered with just yes or no)\begin{itemize}
    \item Yes
    \item No
\end{itemize}

\noindent The language used to describe the activity contains words that are hard for a child aged 5-8 to understand  (e.g. longer than 3-4 syllable words, technical jargon)
\begin{itemize}
    \item Strongly Agree
    \item Somewhat Agree
    \item Neutral
    \item Somewhat Disagree
    \item Strongly Disagree
\end{itemize}

\noindent The activity prompt contains words that a child age 5-8 likely hasn't been exposed to
\begin{itemize}
    \item Strongly Agree
    \item Somewhat Agree
    \item Neutral
    \item Somewhat Disagree
    \item Strongly Disagree
\end{itemize}

\noindent If you put "Somewhat agree or Strongly Agree" to the question above, give a short explanation of what specific words/topics a child likely hasn't been exposed to

\noindent The language used to describe the activity contains complex or long sentences
\begin{itemize}
    \item Strongly Agree
    \item Somewhat Agree
    \item Neutral
    \item Somewhat Disagree
    \item Strongly Disagree
\end{itemize}

\noindent Now you will read a few generated parent conversation starters. These questions are meant to help parents start meaningful discussion with their child about the show they watched and social-emotional themes.

\textbf{\noindent Prompt for the parent to discuss with the child: } [PARENT\_ACTIVITY]

\noindent This question is relevant to the social emotional learning moment: 

\noindent [GPT\_SKILL\_EXPLANATION]. In particular, rate your agreement to the statement: the prompt for the parent mentions/refers to this moment in the show

\begin{itemize}
    \item Strongly Agree
    \item Somewhat Agree
    \item Neutral
    \item Somewhat Disagree
    \item Strongly Disagree
\end{itemize}

\noindent [SKILL\_DESCRIPTION], [SKILL\_DEFINITION], [SKILL\_EXAMPLES], 
The question is related to the social emotional skill explained.
\begin{itemize}
    \item Strongly Agree
    \item Somewhat Agree
    \item Neutral
    \item Somewhat Disagree
    \item Strongly Disagree
\end{itemize}

\noindent The question is likely to foster meaningful dialogue between a parent and child.
\begin{itemize}
    \item Strongly Agree
    \item Somewhat Agree
    \item Neutral
    \item Somewhat Disagree
    \item Strongly Disagree
\end{itemize}

\section{Surveys and Interview Questions} \label{surveys_interviews}
\subsection{Pre-Study Survey}
\begin{itemize}
    \item Q1: What is your child's age?
    \item Q2: What is your age?
    \item Q3: What is your gender?
    \item Q4: What is your child/s gender?
    \item Q5: What do you think social-emotional learning is?
    \item Q6: How are you involved in your child’s social-emotional learning?
    \item Q7: Does your child have a social-emotional learning program at their school?
    \begin{itemize}
        \item Yes
        \item No
        \item Unsure
    \end{itemize}
    \item Q8: How often do you have conversations with your child about the media they consume when you are not around (e.g., YouTube, games, TV, books, etc.)?
    \begin{itemize}
        \item Never
        \item Rarely
        \item Sometimes
        \item Often
        \item Always
    \end{itemize}
    \item Q9: If you have these conversations what do they conversations typically look like? Do you enjoy these conversations?
    \item Q10: How often do you consume media with your child (e.g., watching videos or reading with them)?
    \begin{itemize}
        \item Never
        \item Rarely
        \item Sometimes
        \item Often
        \item Always
    \end{itemize}
    \item Q11: Why do/don't you consume media with your child?
    \item Q12: What types of media do you consume with your child?

\begin{itemize}
    \item TV shows
    \item Apps like YouTube, Instagram, TikTok, etc.
    \item Movies
    \item books
    \item Games
    \item Audiobooks
    \item Podcasts
    \item Other: 
\end{itemize}

    \item Q13: What types of media does your child consume on their own?

\begin{itemize}
    \item TV shows
    \item Apps like YouTube, Instagram, TikTok, etc.
    \item Movies
    \item books
    \item Games
    \item Audiobooks
    \item Podcasts
    \item Other:
\end{itemize}
\item Q14: “Some children spend time watching programmes or videos. We are interested in how common this is for young children. Thinking back over the past two weeks, please indicate how often your child has…”
\begin{itemize}
    \item Watched programmes or videos on a TV/computer/tablet/smartphone with you?
    \item Watched programmes or videos on a TV/computer/tablet/smartphone with someone else?
    \item Watched programmes or videos on a TV/computer/tablet/smartphone on their own?
\end{itemize}
\end{itemize}

\newpage
\subsection{Semi-Structured Interview Questions}
Child question:
\begin{itemize}
    \item Could you tell me what you (liked/disliked) about what we did today?
\end{itemize}

First impressions
\begin{itemize}
    \item What information on this page did you like to see?
    \item What information did you dislike (or could be improved)?
    \item What information is \textit{most} relevant/engaging to you as a parent?
\end{itemize}

\noindent Parent-child Conversation
\begin{itemize}
    \item Imagine you’ve just read this page, and now you’re going to have a conversation with your child. What would talk to them about?
    \item What information on the page influenced the conversation you would have with your child?
    \item What information is missing or not useful for having conversation?
\end{itemize}

\noindent Integration with Daily Life
\begin{itemize}
    \item Imagine you had this tool installed to use at home. Do you think you and your child would use it in daily life? How/Why not?
\end{itemize}

\noindent Social-Emotional Learning

\begin{itemize}
    \item Let’s say this tool is integrated in every video your child watches.
    \begin{itemize}
        \item Do you think the child activity would influence your child’s social emotional learning? How/Why not?
        \item Do you think the parent dashboard would influence your child’s social emotional learning? How/Why not?
    \end{itemize}
    \item In what ways do you see evidence for SEL in your child's response / in what ways did you not?
\end{itemize}

\section{Codebook} \label{codebook}

\onecolumn
\footnotesize

\onecolumn
\footnotesize
\begin{table}[h]
\centering
\begin{tabular}{|p{3cm}|p{0.5cm}|p{3.5cm}|p{7.5cm}|}
\hline
\textbf{Theme} & \textbf{ID} & \textbf{Code} & \textbf{Definition} \\ \hline
\multirow{11}{3cm}{Enhancing Children's Active SEL Learning Practice} 
& 01 & Make media more meaningful and reflective & Parents mention that they would perceive their child consuming media as more meaningful or enriching. Parents feel that their child is getting something out of the video and that the app prevents mindless scrolling and encourages active reflection. \\ \cline{2-4} 
& 03 & Connect to child's life & The tool helps to connect concepts in shows to the child's own life and behaviors. \\ \cline{2-4} 
& 08 & Multiple modes of learning & The tools offer many different ways to learn. \\ \cline{2-4} 
& 04 & Encourages child learning practice & The tool provides the child with extra learning practice. \\ \cline{2-4} 
& 31 & Need more frequent practice for effectiveness & Mention that the tool would be more effective after repeated/routine usage. \\ \cline{2-4} 
& 05 & Sustained child's engagement & Mention that the tool successfully sustained the child's engagement. \\ \cline{2-4} 
& 30 & Better ways to sustain child's engagement & Parents provide suggestions to sustain the child's engagement, for example with gamification. \\ \cline{2-4} 
& 07 & Similar to schoolwork & The tool is similar to homework that is provided to children in school. \\ \cline{2-4} 
& 20 & Allows parent to assess child needs & The tool provides the parent with context to assess the child's social-emotional needs. \\ \cline{2-4} 
& 15 & Extracts important messages/morals/lessons & The tool highlights the most important parts of the story, such as the theme, moral, or lesson. \\ \hline

\multirow{13}{3cm}{Providing Parental Learning and Scaffolding Conversations} 
& 19 & Parent driving home the message & Mention that the tool provides scaffolding for the parent to ultimately drive home the message/lesson. \\ \cline{2-4} 
& 22 & Share life lessons/parent experiences & Parents would share life lessons or personal experiences to teach their child. \\ \cline{2-4} 
& 12 & Encourages parent learning & Point out that the tool is not only helpful for child learning, but also parent learning. \\ \cline{2-4} 
& 21 & Provide evaluation to child & Want expert feedback on how the kid's response to the activity was. \\ \cline{2-4} 
& 10 & Minimizes parent effort & Mention that the tool provides a lighter lift for parents in starting conversation. \\ \cline{2-4} 
& 26 & Information overwhelming & Information was too overwhelming, or too much information was provided. \\ \cline{2-4} 
& 32 & Provides touchpoints for the parent & The tool provides appropriate scaffolding and touchpoints for parents to start a conversation or do teaching. \\ \cline{2-4} 
& 24 & Would talk about child activity & Parents would talk to their child about the activity they did. \\ \cline{2-4} 
& 16 & Enjoy seeing child artifact & Parents particularly enjoy seeing the artifact their child created (drawing, recording, or video). \\ 
& 27 & Behavior different around parent & Since the tool is used independently, children might behave differently than when they know their parent is around. \\ \cline{2-4} 
& 17 & Desire more context during activity & Want more context of the kid when doing activity, for example always providing a video. \\ \cline{2-4} 
& 13 & Summary helpful & Summary is particularly useful for the parent when they did not watch the show. \\ \cline{2-4} 
& 25 & Would talk about video & Parents would talk to their child about the video that they watched. \\ \hline
\end{tabular}
\caption{Codes and Definitions for Themes}
\end{table}

\end{document}